\documentclass[preprint,3p,twocolumn]{elsarticle}
\usepackage{color}

\usepackage{graphicx}
\usepackage{amsmath}
\usepackage{amssymb}
\usepackage{verbatim}
\usepackage{wasysym}
\usepackage{url}
\graphicspath{{Figures/}}
\usepackage{booktabs}
\usepackage{makecell} 
\usepackage[separate-uncertainty=true, output-decimal-marker={.}, per-mode=symbol,range-phrase=-]{siunitx} 
\DeclareSIUnit{\wtpercent}{wt\percent}
\usepackage{multirow}
\usepackage{pifont}
\newcommand{\thickcross}{\ding{54}}%

\biboptions{comma,square,sort&compress}

\journal{Acta Materialia, accepted for publication}

\begin{document}
\begin{frontmatter}
 \title{Multiscale prediction of microstructure length scales in metallic alloy casting}
 \author{B.~Bellon$^{1,2}$}
 \author{A.K.~Boukellal$^{1}$}
 \author{T.~Isensee$^{1,2}$}
 \author{O.M.~Wellborn$^{3}$}
 \author{K.P.~Trumble$^{3}$}
 \author{M.J.M.~Krane$^{3}$}
 \author{M.S.~Titus$^{3}$}
 \author{D.~Tourret$^{1}$}
 \author{J.~LLorca$^{1,2*}$}
 \address{$^1$ IMDEA Materials Institute, C/ Eric Kandel 2, 28906, Getafe, Madrid, Spain.}
 \address{$^2$ Department of Materials Science, Polytechnic University of Madrid/Universidad Polit\'ecnica de Madrid, E.T.S. de Ingenieros de Caminos. 28040 - Madrid, Spain.}
\address{$^3$Purdue Center for Metal Casting Research, School of Materials Engineering, Purdue University, West Lafayette, Indiana, USA.}
 \cortext[cor1]{Corresponding author; Email address: javier.llorca@imdea.org}
\begin{abstract}
Microstructural length scales, such as dendritic spacings in cast metallic alloys, play an essential role in the properties of structural components. Therefore, quantitative prediction of  such length scales through simulations is  important to design novel alloys and optimize processing conditions through integrated computational materials engineering (ICME).
Thus far, quantitative comparisons between experiments and simulations of primary dendrite arms spacings (PDAS) selection in metallic alloys have been mainly limited to directional solidification of thin samples and quantitative phase-field simulations of dilute alloys. 
In this article, we combine casting experiments and quantitative simulations to present a novel multiscale modeling approach to predict local primary dendritic spacings in metallic alloys solidified in conditions relevant to industrial casting processes. To this end,  primary dendritic spacings were measured in instrumented casting experiments in Al-Cu alloys containing 1\,wt\% and 4\,wt\% of Cu, and they were compared to spacing stability ranges and average spacings in dendritic arrays simulated using phase-field (PF) and dendritic needle network (DNN) models. It is first shown  that PF and DNN lead to similar results for the Al-1\,wt\%Cu alloy, using a dendrite tip selection constant calculated with PF in the DNN simulations. PF simulations cannot achieve quantitative predictions for the Al-4\,wt\%Cu alloy because they are too computationally demanding  due to the large separation of scale between tip radius and diffusion length, a characteristic feature of non-dilute alloys. Nevertheless, the results of DNN simulations for non-dilute Al-Cu alloys are in overall good agreement with our experimental results as well as with those of an extensive literature review.  Simulations consistently suggest a widening of the PDAS stability range with a decrease of the temperature gradient as the microstructure goes from cellular-dendrites to well-developed hierarchical dendrites.

\end{abstract}
\end{frontmatter}
%

\section{Introduction}
\label{sec_intro}

Solidification is the initial processing step for most metallic alloys. Microstructural features that develop during solidification play a key role in the properties, performance and lifetime of cast metal parts. 
Even when the microstructure later evolves through additional complex thermomechanical processing steps, the starting point is the microstructure that emerges from the liquid state. 
Dendrites, which are branched structures with primary, secondary and higher-order branches, are among the most common microstructural features in cast alloys \citep{flemings1974solidification,Kurz2019, Trivedi1994}. 
Dendritic microstructures exhibit several important characteristic length scales, such as dendrite tip radius or dendrite arm spacings.
These spacings directly affect the mechanical properties of individual grains \cite{Kurz2019,flemings1974solidification}, in particular ultimate tensile strength and yield strength (see, e.g., \cite{quaresma2000correlation,osorio2002modeling,osorio2006effect}).
They also set the scale for solute (micro)segregation \cite{Kurz2019,flemings1974solidification,beckermann2002modelling}, which has a strong effect on the potential appearance of defects (e.g. freckles) or secondary phases (e.g. precipitates or eutectics), as well as on electrochemical properties, such as corrosion resistance (see, e.g., \cite{osorio2005effect,osorio2006effect,osorio2007roles}).
Primary dendrite arm spacings (PDAS) also play a key role in the permeability of semi-solid microstructures \cite{nasser1985flow,poirier1987permeability,ganesan1992permeability,santos2005permeability,takaki2019permeability}, which has major influence on the appearance of critical defects, such as hot tears.

During growth of primary dendrites, the selection of dendrite tip radius $\rho$ is unique for a given set of solidification conditions (namely, solute concentration $c_\infty$, solidification velocity $V$, and thermal gradient $G$) and follows the microscopic solvability theory \cite{langer1980instabilities, ben1984pattern, benamar1993theory, brener1993needle, karma2000three}. 

The selection of PDAS, on the other hand, is not unique.
A given set of parameters may yield a wide distribution of spacings~\cite{somboonsuk1985dynamical, trivedi1985pattern, Kurz1981, lu1992numerical, hunt1996numerical, warren1993prediction, HT1994}, even though a similar processing history tends to lead to narrowly distributed spacings \cite{HT1994, weidong1993primary}.
Over the years, many studies have focused on the link between the solidification conditions and the morphological characteristics of dendrites (see e.g. \cite{Trivedi1994a, Kurz1981, HT1994, KurzFisher} among others). Most models for the PDAS rely on phenomenological relations linking the primary spacing $\lambda$ to processing parameters in the form of power laws such as \citep{Trivedi1994a}:
\begin{equation}
    \lambda=KG^{-a}V^{-b}
\end{equation}\label{eq:generic_power_law_pdas}
\noindent where the exponent $a$ is often close to 0.5, $b$ typically varies between 0.25 and 0.5, and $K$ is a prefactor that depends on alloy parameters, such as its composition and phase diagram features (e.g. solute partition coefficient) \cite{Hunt1979,Kurz1981,HT1994,KurzFisher}. 
However, experimental evidence indicates that a wide range of PDAS can be obtained under similar solidification conditions. 
The lower limit of the PDAS stability range, $\lambda_{min}$, is linked to an elimination instability \citep{HT1994,Boettinger2000, Provatas2010, tourret2015oscillatory,tourret2015growth}, while the upper stability limit, $\lambda_{max}$, stems from a branching instability.  The determination of this range is not, however, trivial considering all the mechanisms involved, including: tip elimination through thermal and solutal interaction, tip splitting, secondary and tertiary arm development, lateral migration, thermo-solutal convection, and defect interaction with the growth front. 

Directional solidification is the standard experimental approach to establish the link between solidification conditions and microstructure in metallic alloys. Particularly, dendrite growth in the Al-Cu system has been studied for many solidification conditions ($G$ and $V$), different Cu concentrations and different experimental set-ups. They include standard Bridgman furnaces in which the solidification velocity is controlled \citep{Geying1987,McCartney1981, Lin1999, Gunduz2002}, or small volumes in which solidification takes place under a given thermal gradient \cite{Okamoto1975,Young1975,Kaneko1977, Su1998, Grange1995}, while other studies used larger molds in which the temperature was measured along the solidification direction to obtain the thermal gradient and the solidification velocity  \citep{Quaresma2000, Rocha2003, Spinelli2004, Eskin2005}. 
The differences in solidification conditions has led to a significant scatter in the PDAS for similar nominal values of the solidification process, as given by the cooling rate $R$ = $GV$, but the overall trend is clearly indicative of the power law behavior from phenomenological scaling laws (see Fig.~I of the Supplementary Material). 


Further analysis of the parameters controlling microstructural development during solidification can be obtained through modeling and simulation. 

Conventional models for macroscopic solidification processes (e.g. casting) rely on conservation equations averaged over multi-phase domains, such that a field corresponding to the average fraction of phase over representative volume elements appears in the equations \cite{rappaz2010numerical, ni1991volume, wang1996equiaxed_1, wang1996equiaxed_2, combeau2009prediction, wu2009modeling_1, wu2009modeling_2}. 
The fraction of phases may be tabulated as a function of temperature using classical solidification paths such as lever rule or Gulliver-Scheil model \cite{KurzFisher}. Alternatively, microstructural length scales (e.g. dendritic spacings) may be introduced in the equations to model the kinetics of solute microsegregation between dendritic branches a bit more accurately \cite{wang1996equiaxed_1, wang1996equiaxed_2, combeau2009prediction, wu2009modeling_1, wu2009modeling_2, tourret2011multiple, tourret2011gas}. These length scales are then typically estimated from phenomenological power laws such as Eq.~\eqref{eq:generic_power_law_pdas}.

Models capable of estimating microstructural length scales require a relatively more precise spatial description, consistent with the scale of the features one aims at predicting.
This makes simulations computationally challenging since they have to combine phenomena across a wide range of scales, from microscopic capillarity at dendritic tips to macroscopic transport of heat and solute in the melt. 
Simulations of dendritic spacings have been performed with a range of models based on cellular automata (CA) \cite{rappaz1993probabilistic, gandin1994coupled, wang2003model}, grain envelopes \cite{steinbach1999three, viardin2017mesoscopic}, dendritic needle networks (DNN) \cite{tourret2013multiscale, tourret2016three, tourret2015three_1, tourret2015three_2} and phase fields (PF) \cite{echebarria2010onset, gurevich2010phase, clarke2017microstructure}.
Approximate --- and computationally efficient --- models (e.g. CA) rely on stringent assumptions and phenomenological laws, for instance directly relating the growth kinetics of dendrite tips to some measure of the local driving force for solidification (e.g. undercooling or solute supersaturation measured at a given distance from the solid-liquid interface). 
The most accurate models (e.g. PF) rely solely on fundamental concepts of diffuse interface thermodynamics \cite{boettinger2002phase, steinbach2009phase} and allow quantitative calculations directly from thermophysical alloy parameters and processing conditions. 
However, they remain limited in length and time scales by their relatively high computational cost.
Hence, a multiscale multi-model bottom-up strategy appears to be a promising route to address the limitations of each models.

Here we specifically focus on two modeling methods, namely PF and DNN, and we show that quantitative PF simulations of dendritic growth in dilute alloys can be quantitatively upscaled and expanded to non-dilute compositions using the multiscale DNN approach.

The PF method is now recognized as the outstanding tool to predict microstructural patterns resulting from nonequilibirum solidification \cite{boettinger2002phase, steinbach2009phase, karma2016atomistic}.
The method allows modeling the evolution of solid-liquid interfaces and complex microstuctural patterns, such as dendrites.
It has been applied to the solidification of Al-Cu alloys, at first qualitatively, e.g. to reproduce scaling laws for secondary dendrite arm spacing in cast samples \cite{ode2001numerical}. 
Then, the development of quantitative models for dilute binary alloys \cite{karma1996phase, karma1998quantitative, karma2001phase, echebarria2004quantitative} has allowed direct quantitative comparisons with directional solidification experiments of dilute Al-Cu thin-samples imaged in situ with X-ray radiography \cite{clarke2017microstructure}.
However, quantitative PF predictions are still constrained by the requirement of an accurate description of the local solid-liquid interface curvature.
This limits the size of the grid elements required in the vicinity of each dendritic tip for primary, secondary, and higher-order side-branches.
This limitation may, to some extent, be addressed through parallelization \cite{shimokawabe2011peta, shibuta2015solidification} and advanced numerics (e.g. adaptive meshing \cite{provatas1998efficient, provatas1999adaptive, greenwood2018quantitative} or implicit time stepping \cite{rosam2008adaptive, bollada2015three}).
However, quantitative PF simulations become challenging, if feasible at all, beyond the dilute alloy limit, as the separation of scale between interfacial capillary length, dendrite tip radius, and solute transport in the bulk becomes extreme.

In order to overcome this limitation, the dendritic needle network (DNN) method was developed to bridge the scale of the microscopic dendrite tip radius, $\rho$, and that of solute transport in the liquid phase, e.g. the diffusion length $l_D=D/V$ with $D$ the solute diffusivity. The DNN model represents a dendritic crystal as a network of parabolic-shaped needles, that approximate the primary stems and higher order branches. The DNN model combines the microscopic solvability theory established at the scale of the tip radius \cite{langer1980instabilities, ben1984pattern, benamar1993theory, brener1993needle} with a solute mass balance established at the ``mesoscale'' between $\rho$ and $l_D$, to calculate the instantaneous growth velocities of individual dendrite tips \cite{tourret2013multiscale, tourret2016three, tourret2019multiscale, isensee2020three}.
The numerical spatial discretization can be taken at the same order as the tip radius, such that the domain size can be considerably larger than in equivalent PF simulation, in particular for concentrated alloys with $\rho\ll l_D$. 
DNN simulations have been verified to reproduce analytical and phase-field predictions for steady-state and transient growth kinetics \cite{tourret2013multiscale, tourret2016three, tourret2020comparing}.
First DNN applications to the prediction of primary dendritic spacings in binary Al alloys, compared to well-controlled thin-sample directional solidification experiments \cite{tourret2015three_iop, tourret2015three_jom, tourret2016three}, have shown promising results.

This investigation is aimed at bridging length scales in the simulation of alloy solidification.
One central objective is to demonstrate that microstructural length scales, such as PDAS, can be computationally predicted for a bulk specimen, under conditions distinctive from a Bridgman type experiment, namely nonuniform $G$, $V$ and non-negligible fluid flow, making it closer and thus more representative of solidification conditions for industrial casting processing.
To do so, we performed casting experiments of binary Al-Cu alloys, and combined simulations at distinct length scales.
First, we show that both PF and DNN can provide reliable predictions of stability range of PDAS for an Al-1\,wt\%Cu alloy.
Then, we use the computationally-efficient DNN model to calculate the PDAS stability range for a more concentrated Al-4\,wt\%Cu alloy. 
We discuss and compare our results with a broad range of experimental literature data for Al-Cu alloys.
Overall, the article shows how multiscale modeling can provide quantitative predictions of important microstructural features (e.g. PDAS) during solidification of non-dilute alloys at experimentally relevant length and time scales using physics-based computational models.

\section{Methods}
\label{sec_methods}

Here, we provide a short description of our experimental and computational methods. 
A more thorough description with full details of equipments, procedures, equations, and algorithms is provided in the joint Supplementary Material document.

\subsection{Experiments and characterization}
\label{sec_meth_exp}

\subsubsection{Casting experiments}

\begin{figure}[!b]
\centering
 	\includegraphics[width=7.7cm]{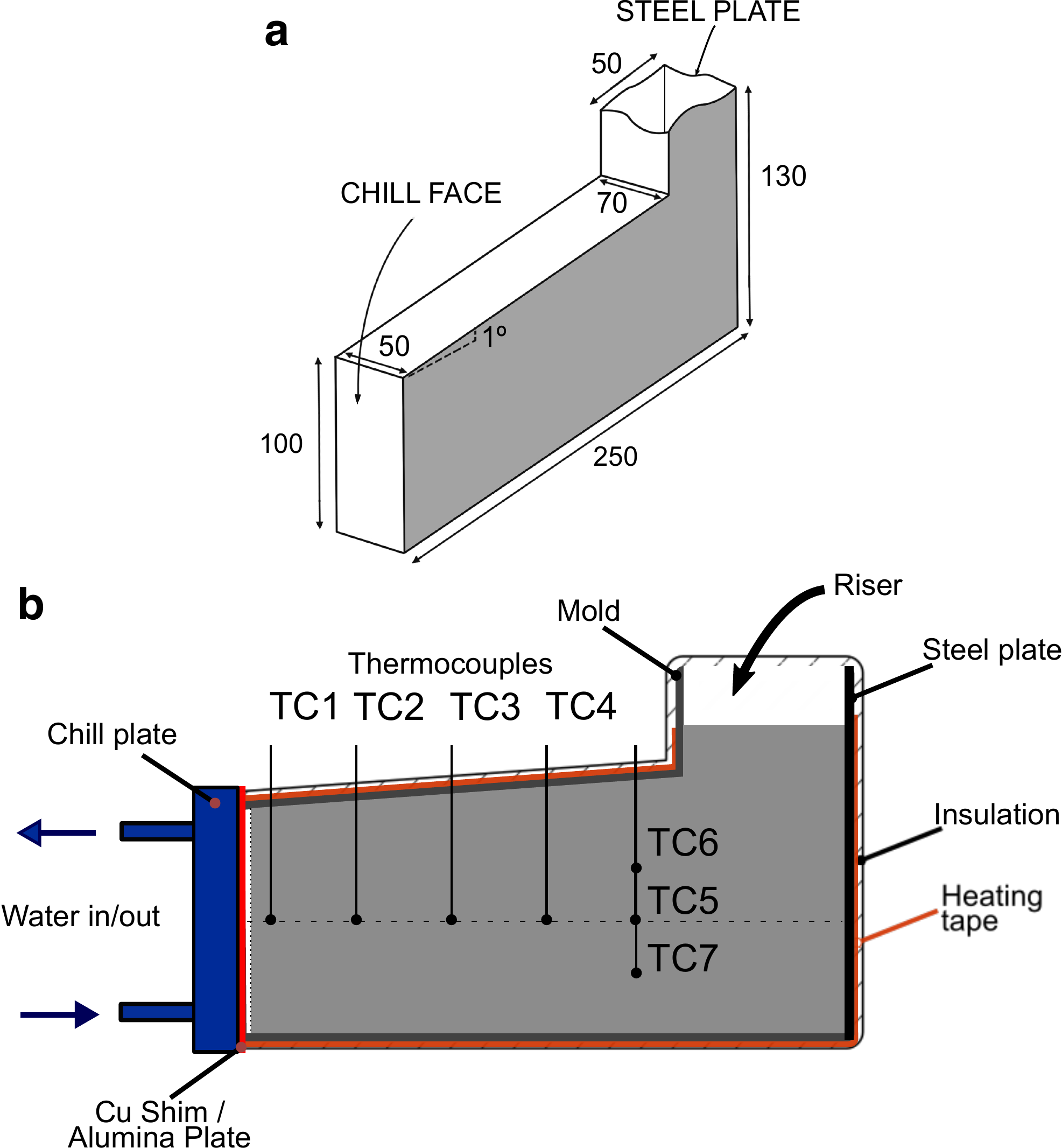}
	\caption{
	Experimental setup for casting experiments: 
	(a) Schematics and dimensions (in mm) of the mold; 
	(b) Mold assembly schematics (not to scale) showing the locations of thermocouples TC1 through TC7.
	}
	\label{fig:mold}
\end{figure}

We performed casting experiments in the facilities of the School of Materials Engineering at Purdue University. Two different Al-Cu alloys with 1 and 4\,wt\%Cu were prepared by melting Al granules and Cu rods in a clay-graphite crucible using an air induction furnace. The alloys were poured into a stainless steel mold, of dimensions shown in Fig.~\ref{fig:mold}a. The mold assembly was wrapped with a heating tape to reach approximately 500$^\circ$C before pouring. Prior to pouring through the riser, the heating tape was disconnected. The exterior of the mold was covered with several layers of an insulating blanket to minimize the heat losses through the mold walls.   

The heat was extracted from the side of the mold by a water-cooled chill plate (Fig. \ref{fig:mold}b), promoting predominantly horizontal directional solidification. For each alloy, we explored different cooling rates by performing two sets of experiments: either (1) with a 0.2~mm thick copper shim between the chill plate and the mold, thus ensuring good thermal contact, or (2) with an additional 5~mm thick alumina plate between the chill plate and the copper shim, thus reducing the heat extraction rate.
The mold was instrumented with 7 K-type thermocouples, shown in Fig.~\ref{fig:mold}b (also see Table~I of the Supplementary Material), mechanically joined and protected from dissolution by high-temperature cement. 

%
\subsubsection{Microstructure characterization}

We sectioned the cast ingot in between the locations of thermocouples and cut out cross-sections of about $30\times30~$mm$^2$. 
Their surface was polished using SiC paper up to 4000 grit, followed by diamond pastes down to 0.25 $\mu$m particle size, and finishing with a solution of colloidal silica (OP-S).  

The polished samples were etched with Kellers reagent and observed under two different optical microscopes with magnification from 6.3$\times$ to 1000$\times$. 
We used the software ImageJ to enhance the image contrast and manually located the apparent centers of dendrites, measuring at least 400 primary dendrites in each specimen. 
The average PDAS was calculated from these locations using a standard Delaunay triangulation Python code. 

We performed electron backscatter diffraction (EBSD) on the cross-sections in order to assess the misorientation between the typical [100] growth direction of Al dendrites and the cutting/observation plane.
Several EBSD scans were combined into grain orientation maps in areas of about $16\times4$~mm$^2$ and analyzed using the MTEX toolbox \cite{hielscher2019denoising}.


\subsubsection{Estimation of local solidification conditions}
\label{sec:meth_local_conditions}

In this article we focus primarily on the columnar dendritic structures observed between the locations of thermocouples TC1 and TC2, i.e. in the early stage of the experiment when the solidification direction is still expected to be aligned with the (horizontal) main heat extraction direction.

We estimated the velocity of isotherms between TC1 and TC2 by dividing the distance between the thermocouples by the times $t_1$ and $t_2$ at which the temperatures from TC1 and TC2 reach a given temperature $T_L^-$ chosen up to 20~K below the alloy liquidus temperature $T_L$ (see Section~\ref{sec_resu_casting} and Fig.~\ref{fig:Temp_time}). 

The local temperature gradient was estimated as an average cooling rate $R$ of the temperatures recorded by TC1 and TC2 between times $t_1$ and $t_2$.
Practically, we performed a linear fit of TC1 and TC2 in that time range to extract their respective cooling rates $R_1$ and $R_2$, and then considered that $R\approx(R_1+R_2)/2$ to calculate $G=R/V$.

Since the estimation of the local velocity of isotherms $V$ was more accurate than that of the temperature gradient $G$ (see section~\ref{sec_resu_casting}), we decided to use $V$ as a fixed input value in both PF and DNN simulations, and to explore different values of $G$, i.e. different cooling rates $R=GV$.

\subsection{Modeling and simulations}
\label{sec_meth_simu}

We used two models already described and used in several previous publications.
Therefore, we only give a brief description of the models and the main simulations procedures.
Nevertheless, a complete description of model equations, algorithms, and an exhaustive list numerical and alloy parameters, together with a brief discussion of the latter, are provided in the joint Supplementary Material.

\subsubsection{Phase-field}
\label{sec_meth_pf}

We used a quantitative three-dimensional (3D) PF model for the directional solidification of a dilute binary alloy in a temperature gradient \cite{karma1996phase, karma1998quantitative, karma2001phase, echebarria2004quantitative} to calculate, in conditions representative of our experiments: (1) the limits of the PDAS stability range $\lambda_{min}$ and $\lambda_{max}$, (2) the dynamically selected spacing in a spatially extended array with several primary dendrites, and (3) the dendrite tip selection parameter $\sigma$.
All PF simulations were performed for the Al-1\,wt\%Cu experiment using one growth velocity $V=160~\mu$m/s, estimated in the experiment with the insulating alumina plate (see Section \ref{sec_resu_casting}), and exploring different temperature gradients $G=10^2$, $10^3$, and $10^4$~K/m

\begin{figure}[!b]
\centering
\includegraphics[width=7cm]{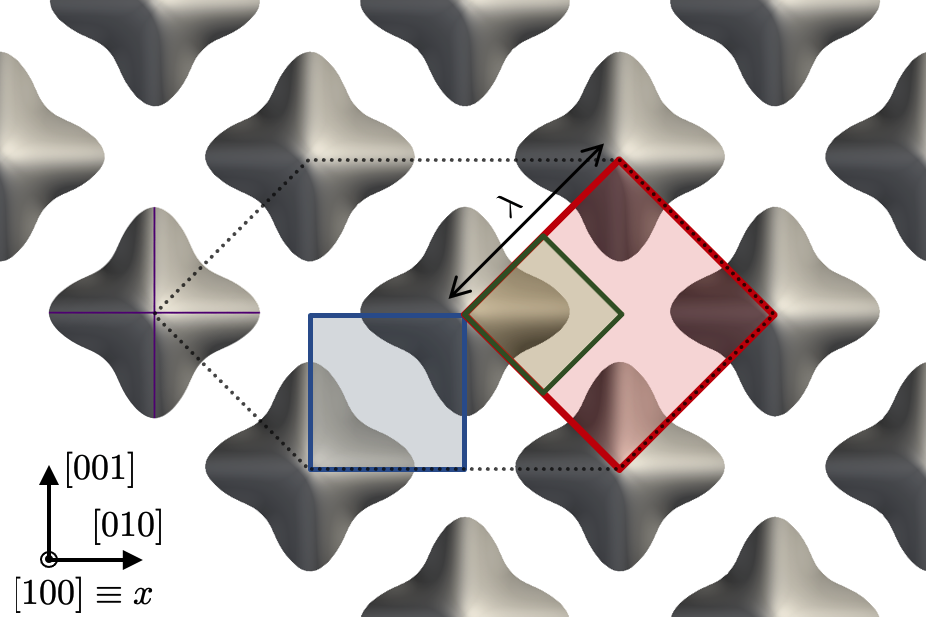}
\caption{
Simulation domains for a quincunx (fcc-like) dendritic pattern of primary spacing $\lambda$, as used in the phase-field (green and red squares) and DNN (blue square) simulations. For clarity only the tip region is shown here (up to $\approx165~\mu$m behind the tip) and not the region at lower temperature, in which side-branching may occur.
\label{sim_sym}
}
\end{figure}

First, we calculated $\lambda_{min}$ and $\lambda_{max}$ in a regular quincunx (fcc-like) dendritic array, illustrated in Fig.~\ref{sim_sym}.
Thus, we simulated one quarter of a dendrite (green square in Fig.~\ref{sim_sym}) with mirror symmetry imposed along all boundaries.
Using a procedure already used in several previous publications \cite{tourret2015oscillatory, pereda2017experimental, song2018thermal} (see Supplementary Material), we progressively increased the size of the domain along $y$ and $z$ until a branching event occurs, thus marking the upper limit of spacing stability $\lambda_{max}$. 
In order to identify the lower stability limit $\lambda_{min}$, we performed simulations with four quarters of dendrites in opposite corners of the simulation box (red square in Fig.~\ref{sim_sym}), and reduce the domain size until one or more primary dendrite is eliminated, thus identifying $\lambda_{min}$.
Using a moving frame following the tip location in the $x$ direction, the size of the domain in $x$ was taken long enough for no-flux boundary conditions to have a negligible effect on the tip growth dynamics (see Supplementary Material).

Then, we performed one simulation of a spatially extended array of dendrites in a thermal gradient $G=10^3$~K/m, using a simulation domain of cross-section $L_y\times L_z  = 512\times512 ~\mu$m$^2$ and length $L_x=1185~\mu$m, starting from a planar interface located at the liquidus temperature.
The simulated time was 200\,s, which was found to be enough to approach an apparent steady-state microstructure. 

We also used the PF results to estimate the dendrite tip selection parameter $\sigma$, which is an input parameter of the DNN model.
To do so, we extracted primary tip radii in simulations of one quarter of dendrite, using least-square fitting of the solid-liquid interface in (010) and (001) planes to a second order parabola (purple lines on the left-side dendrite of Fig.~\ref{sim_sym}) using the method presented in Ref.~\cite{clarke2017microstructure} (see Supplementary Material).
We calculated $\sigma$ for a handful of steady-state dendrites at different spacings in order to verify its independence to the PDAS in the simulations of one quarter of dendrite at $G=10^2~$K/m.
We selected the lowest of the explored temperature gradients as it is well within the well-developed dendritic regime, i.e. far from the cell-to-dendrite transition regime. 

\subsubsection{Dendritic Needle Network}
\label{sec_dnn}

We used a multiscale 3D DNN model for directional solidification already presented elsewhere~\cite{tourret2013multiscale, tourret2016three, tourret2019multiscale} in order to calculate, for both Al-1\,wt\%Cu and Al-4\,wt\%Cu alloys: (1) the PDAS stability limits $\lambda_{min}$ and $\lambda_{max}$ and (2) dynamically selected spacings in a spatially extended dendritic array.

For each alloy, we used one growth velocity estimated between thermocouples TC1 and TC2 in the experiments with an insulating alumina plate, namely $V=160~\mu$m/s (Al-1\,wt\%Cu) and $V=200~\mu$m/s (Al-4\,wt\%Cu).
We scanned a range of temperature gradient $G=10^2$, $10^3$, and $10^4$~K/m for $c_\infty=1$\,wt\%Cu and $G=5\times10^2$, $5\times10^3$, and $5\times10^4$~K/m for $c_\infty=4$\,wt\%Cu.
The value of the tip selection parameter, $\sigma=0.063$, was estimated from phase-field simulation results.

To estimate both $\lambda_{min}$ and $\lambda_{max}$, DNN simulations were initialized with two primary dendrites in opposite corners of the simulation domain (blue square in Fig.~\ref{sim_sym}).
Simulations were performed for different domain sizes $L_y=L_z=\lambda/\sqrt{2}$ in order to identify spacings leading to the elimination of one of the two primary dendrite (when $\lambda<\lambda_{min}$) or to the emergence of a tertiary branch as a new primary dendrite (when $\lambda>\lambda_{max}$).
Similarly as in PF simulations, we use a moving frame in the $x$ direction and the length of the domain in $x$ was chosen long enough for boundary conditions to have a negligible effect (see Supplementary Material).

Additionally, we performed two simulations of spatially extended dendritic arrays to determine a dynamically selected average PDAS $\langle \lambda \rangle$. 
For the Al-1\,wt\%Cu alloy, we used $G=10^3$~K/m and $V=160~\mu$m/s in a domain of size $L_y=L_z=1138~\mu$m, and for the Al-4\,wt\%Cu alloy, we used $G=5\times10^2$~K/m and $V=200~\mu$m/s in a domain of size $L_y=L_z=692~\mu$m, both with a length $L_x=1280~\mu$m.
The domain was initialized with respectively 1849 and 576 uniformly distributed primary needles evenly spaced in $y$ and $z$, in order to approximate a planar interface located at the alloy liquidus temperature.
The simulated time were $200$\,s and $100$\,s, respectively, which was found sufficient to approach a steady microstructure with no further primary dendrite elimination.

\section{Results}
\label{sec_results}

\subsection{Casting}
\label{sec_resu_casting}

\begin{figure}[!b]
	\centering
	\includegraphics[width=7.2cm]{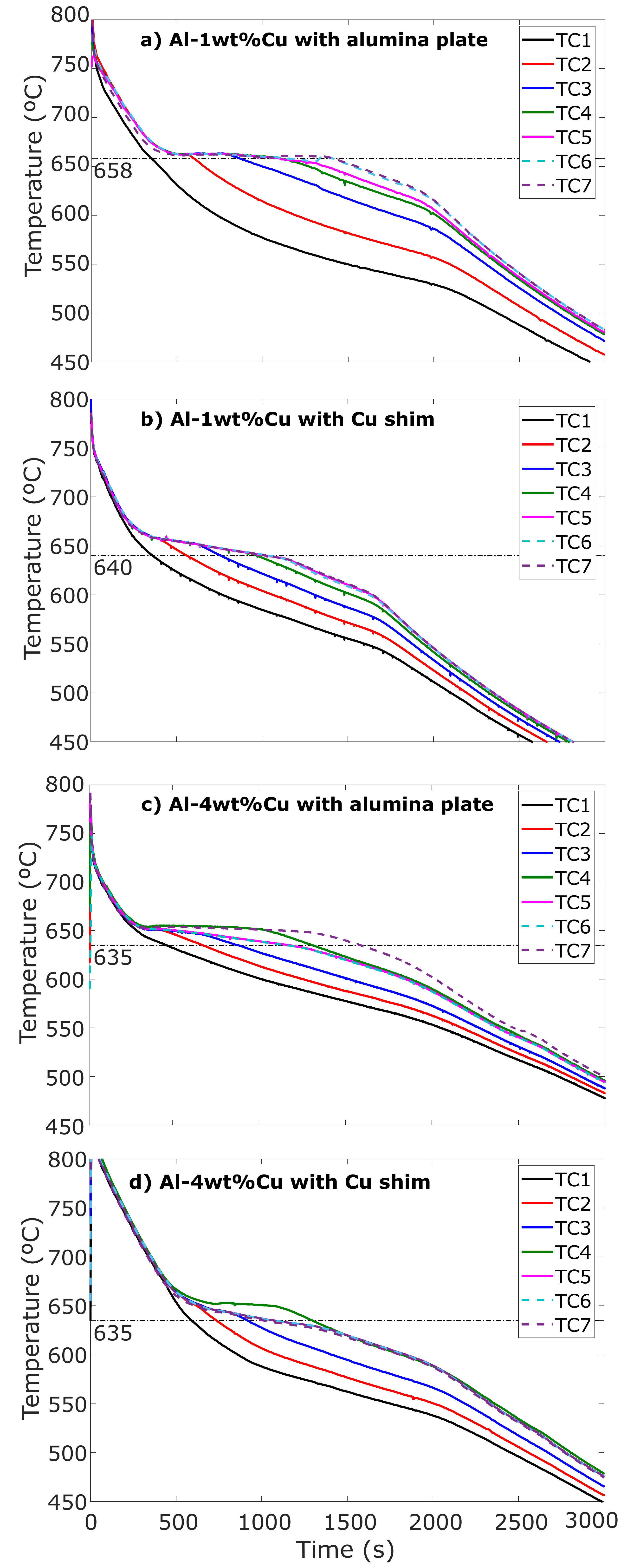}
	\caption{Temperatures measured by the 7 thermocouples during the casting experiments. Al-1\,wt\%Cu: (a) with the alumina plate between the mold and the chill plate and (b) with direct contact between the mold and the chill plate. Al-4\,wt\%Cu: (c) with the alumina plate between the mold and the chill plate and (d) with direct contact between the mold and the chill plate.
	Horizontal dash-dot lines mark the temperature at which the velocity of isotherms was estimated.
	}
	\label{fig:Temp_time}
\end{figure}

The temperature-time curves recorded by the 7 thermocouples are plotted in Fig.~\ref{fig:Temp_time} for the four experiments. 
All curves exhibit a plateau or a notable inflection below the liquidus temperature, indicative of the release of latent heat (recalescence) during solidification.

Readings from thermocouples close to the cooling surface (i.e. TC1, TC2, and TC3) are nearly parallel to one another for a notable range of temperature below liquidus temperature $T_L$.
Thus, the velocity of isotherms at $T=T_L^-\lesssim T_L$ (dash-dot lines in Fig.~\ref{fig:Temp_time}) --- calculated as $V\approx d_{12}/(t_2-t_1)$, with $d_{12}$ the distance between TC1 and TC2, and $t_1$ and $t_2$ the times at which TC1 and TC2 respectively cross the temperature $T_L^-$ --- provides a reasonably accurate estimation of the average solidification front velocity between TC1 and TC2.

The local temperature gradient is less straightforward to obtain, as it depends significantly on the method used to calculate it. 
As explained in Section~\ref{sec:meth_local_conditions}, it is calculated from the time-averaged slopes of TC1 and TC2 between $t_1$ and $t_2$.

Resulting solidification conditions between TC1 and TC2 are summarized in Table~\ref{tab_gv}.
In the simulations, we specifically focus on the two experiments performed with an insulating alumina plate, i.e. cases (a) and (c) in Fig.~\ref{fig:Temp_time} and Table~\ref{tab_gv}.

\begin{table}[h]
\begin{center}
\begin{tabular}{cccc}
\hline
Case & V ($\mu$m/s) & G (K/mm) & R (K/s) \\
\hline
\hline
(a) & $0.16 \pm 0.01$ & $0.91 \pm 0.29$ & $0.15 \pm 0.04$ \\
(b) & $0.23 \pm 0.02$ & $0.50 \pm 0.01$ & $0.12 \pm 0.03$ \\
(c) & $0.19 \pm 0.01$ & $0.37 \pm 0.12$ & $0.07 \pm 0.02$ \\
(d) & $0.28 \pm 0.02$ & $0.50 \pm 0.20$ & $0.14 \pm 0.04$ \\
\hline
\end{tabular}
	\caption{Solidification velocity $V$, temperature gradient $G$, and cooling rate $R$ between thermocouples TC1 and TC2: Al-1\,wt\%Cu alloy (a) with and (b) without alumina plate between mold and chill plate; Al-4\,wt\%Cu alloy (c) with and (d) without alumina plate between mold and chill plate.
	}
\label{tab_gv}
\end{center}
\end{table}

\subsection{Microstructure characterization }
\label{sec_resu_charac}

\begin{figure}[!t]
	\centering
	\includegraphics[width=7cm]{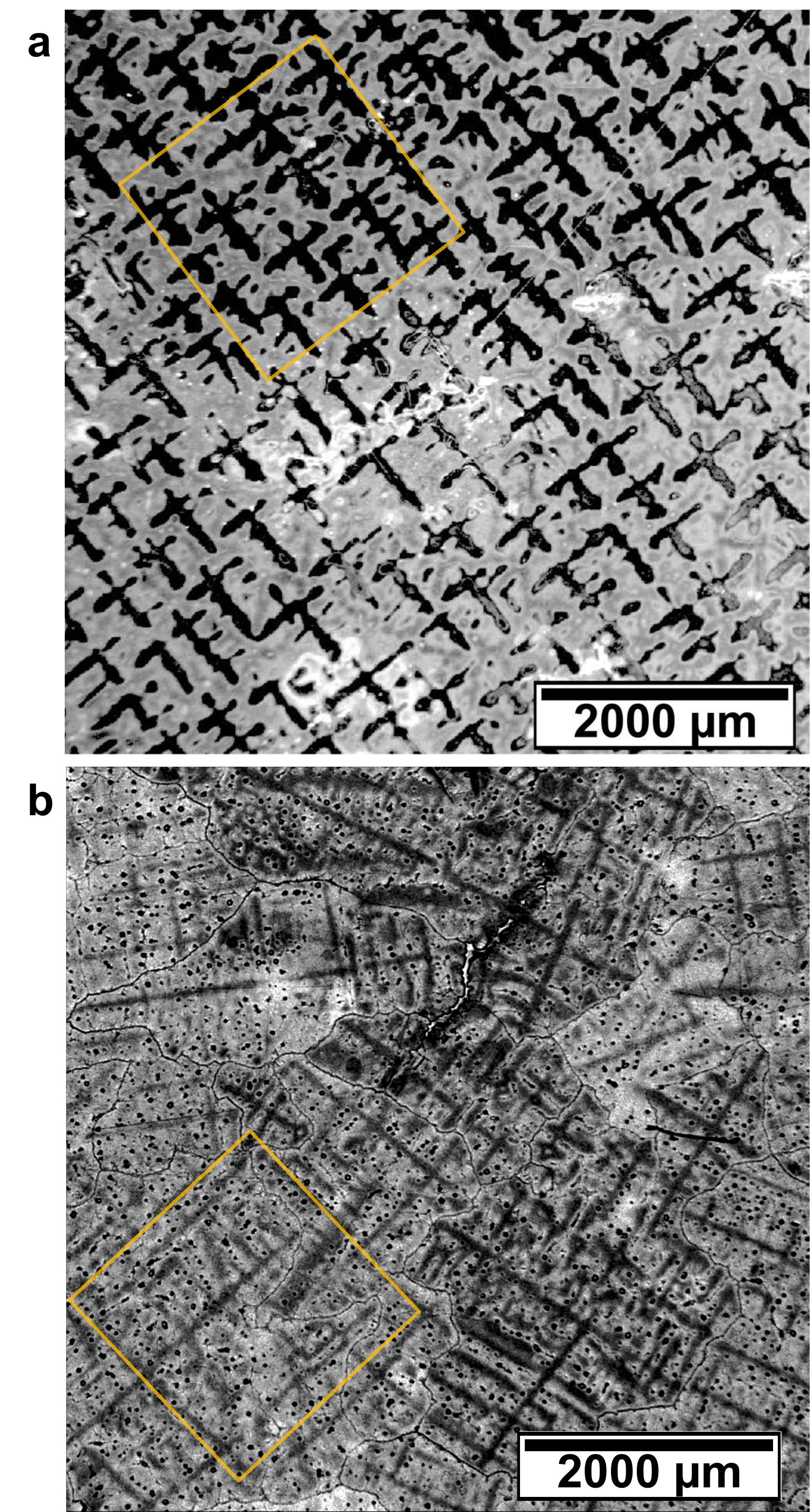}
	\caption{
	Micrographs of the cross-section of the (a) Al-1\,wt\% Cu and (b) Al-4\,wt\% Cu alloy solidified using the alumina plate between the mold and the chill plate at the midpoint between thermocouples 1 and 2. 
	Orange squares show the regions compared to simulation results in Fig.~\ref{fig:simu_spac_array}.
	\label{fig:dendrites-alumina}
	}
\end{figure}

Cross-section micrographs between TC1 and TC2 appear in Fig.~\ref{fig:dendrites-alumina} for the two alloys solidified using the alumina plate between mold and chill plate.

\begin{figure}[!t]
	\centering
	\includegraphics[width=8cm]{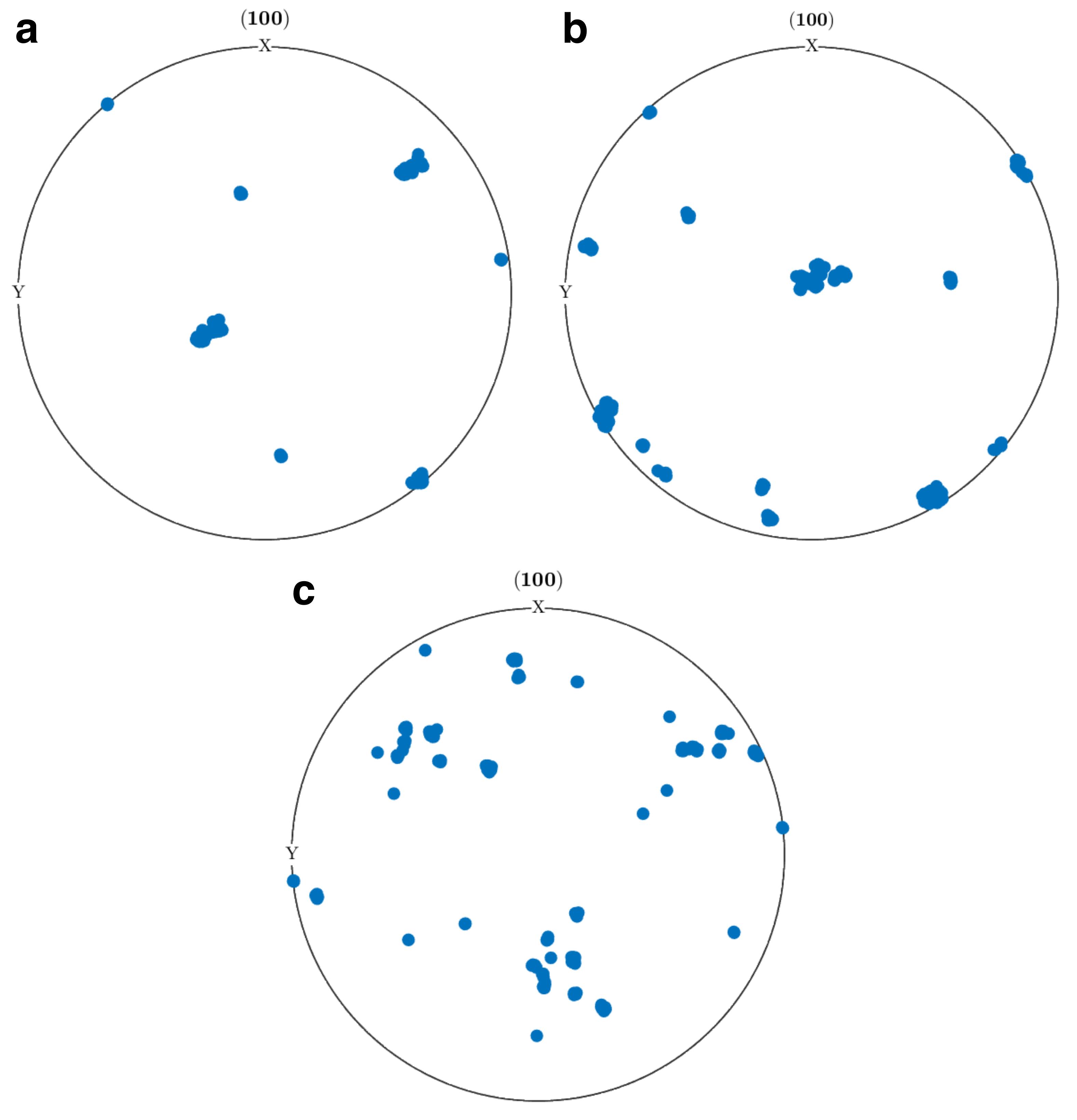}
	\caption{
	EBSD [100] pole figures for the Al-1\,wt\% Cu alloy (a) between TC1 and TC2 and (b) between TC3 and TC4 as well as (c) for the Al-4\,wt\% Cu between TC3 and TC4.
	The scatter in a and c shows the lack of overall alignment of the [100] dendrite growth direction with the observation plane, in contrast to panel b showing a higher density of grains growing along the principal temperature gradient direction.
	}
	\label{fig:pole100}
\end{figure}

Figure \ref{fig:pole100} shows the [100] pole figures from EBSD analysis of the dendrite orientation with respect to the cutting plane in 16$\times$4~mm$^2$ cross-section areas.  
We assumed that dendrites grow along their [100] crystalline direction, since they are in a well-developed dendritic regime, far from a cellular-to-dendritic transition regime in which a deviation may exist between [100] and growth direction \cite{akamatsu1997similarity, deschamps2008growth}.

For the Al-1\,wt\%Cu alloy, between TC3 and TC4 the cutting plane was nearly normal to the [100] direction (Fig.~\ref{fig:pole100}b) but closer to the chill plate, e.g. between TC1 and TC2, the misorientation was more notable (Fig.~\ref{fig:pole100}a). 
In the Al-4\,wt\%Cu alloy (Fig.~\ref{fig:dendrites-alumina}b), most cross-sections presented several grains of different orientations with no remarkable preferential orientation (Fig.~\ref{fig:pole100}c).

\begin{figure}[!t]
	\centering
	\includegraphics[width=7cm]{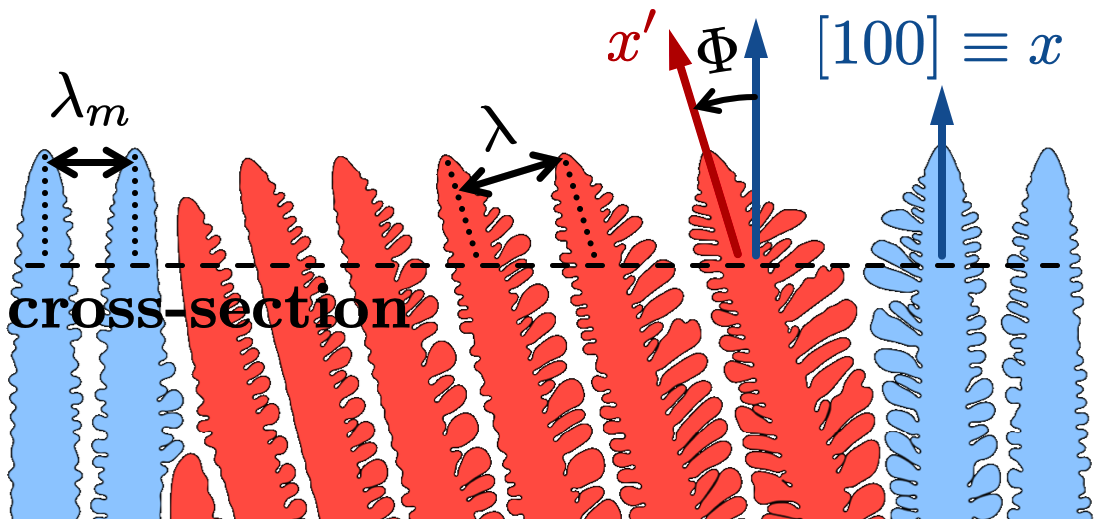}
	\caption{
	Schematic for the correction of the dendrite spacing from the dendrite orientation, where $\lambda_m$ is the PDAS measured on the cross-section and $\lambda$ is the actual PDAS. 
	}
	\label{fig:deviated_dendrites}
\end{figure}

We used the Euler angle $\Phi$ to estimate the deviation between the apparent spacing in the cross-sections $\lambda_m$ and the actual spacing normal to the growth direction $\lambda \approx \cos(\Phi)\lambda_{m}$ (Fig.~\ref{fig:deviated_dendrites}).
The angle $\Phi$, negligible for most grains between TC3 and TC4 in the Al-1\,wt\%Cu alloy, reached up to 40$^\circ$ between TC1 and TC2 and in the the Al-4\,wt\%Cu alloy, thus leading to $\lambda$ up to 25\% lower than $\lambda_m$. 
Hence, for all samples, we estimated $\lambda\approx 0.9 \lambda_m$ as an average approximate angle correction.

The average measured and corrected PDAS are plotted as a function of the cooling rate $R$ in Fig.~\ref{fig:PDAS_Al1CU} (Al-1\,wt\%Cu) and Fig.~\ref{fig:PDAS_Al4CU} (Al-4\,wt\%Cu), together with literature data for relevant composition ranges (see graph legend), and a dotted line illustrating a $R^{-0.5}$ slope. 

Even though PDAS were slightly lower when the mold was in direct contact with the chill plate, this effect was relatively small because the two achieved cooling rates remained of the same order (see Fig.~\ref{fig:Temp_time}), which was not sufficient to yield a significant change microstructural length scale.
Therefore, in Figs~\ref{fig:PDAS_Al1CU} and \ref{fig:PDAS_Al4CU}, we did not distinguish between experiments with and without the insulating alumina plate, since the resulting scatter is negligible compared to the overall cloud of points. 

\subsection{Simulations}
\label{sec_resu_simu}

The first quantitative results extracted from the PF simulation are the dendrite tip radii and the corresponding tip selection parameter $\sigma$. 
Results from simulations of the Al-1\,wt\%Cu alloy at $V=160~\mu$m/s and $G=10^2~$K/m for different spacings $\lambda$ appear in Table~\ref{tab:rad_sigma}.
The average radius remains within $\rho = 8.96 \pm0.13~\mu$m, with less than 4\% difference between radii measured in the $y$ and $z$ directions.
The resulting tip selection parameter, accounting for the tip temperature, i.e. for the temperature-dependent capillarity length, is within $\sigma = 0.06 \pm 0.002$.
This compares well with linear solvability theory, which gives $\sigma\approx0.06$ for a one-sided model (see, e.g., Fig.~2 of Ref.~\cite{barbieri1989predictions} with $\alpha=15\epsilon_4=0.18$ and Eq.~(4.3) therein with $\mu=0$). 
In the DNN simulations we used $\sigma=0.063$, which was extracted from preliminary PF simulations performed before we obtained all PF results.
However, we expect that such small difference had little influence on any of our results and conclusions.

\begin{table}[!t]
\centering
\caption{
Fitted tip radii and selection parameters from the PF simulations of the Al-1\,wt\%Cu alloy at $V=160~\mu$m/s and $G=10^2~$K/m.
}
\label{tab:rad_sigma}
\begin{tabular}{c c c}
\toprule
$\lambda$ ($\mu$m) & $\rho$ ($\mu$m) & $\sigma$ \\ 
\midrule
336 &  8.94 & 0.0605 \\
387 &  8.93 & 0.0606 \\
438 &  8.83 & 0.0619 \\
490 &  9.09 & 0.0585 \\
505 &  9.06 & 0.0589 \\
521 &  9.02 & 0.0593 \\ 
\bottomrule
\end{tabular}%
\end{table}

\begin{figure}[!t]
	\centering
	\includegraphics[width=7.8cm]{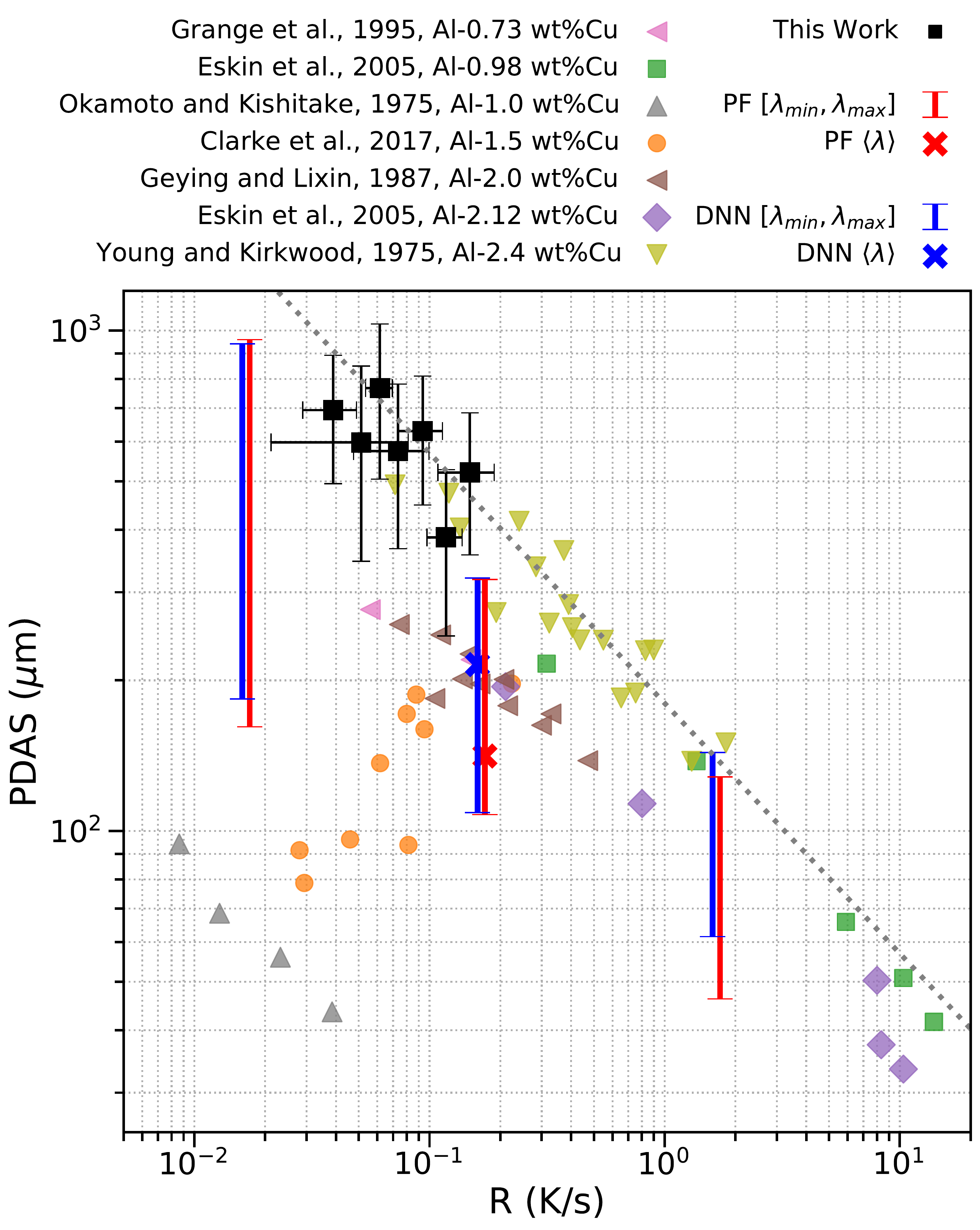}
	\caption{
	Comparison of PDAS as a function of cooling rate for the Al-1\,wt\%Cu alloy in our experiments (both cooling conditions), together with literature results, as well as PF (red) and DNN (blue) predictions of dendritic stability range (error bars) and average spacing in spatially extended dendritic arrays (\thickcross). 
	PF and DNN simulations are performed at the same $(G,V)$ and hence $R$, but PF data points in the plot are slightly shifted to the right, for the sake of readability.
	}
	\label{fig:PDAS_Al1CU}
\end{figure}

\begin{figure}[!t]
	\centering
	\includegraphics[width=7.8cm]{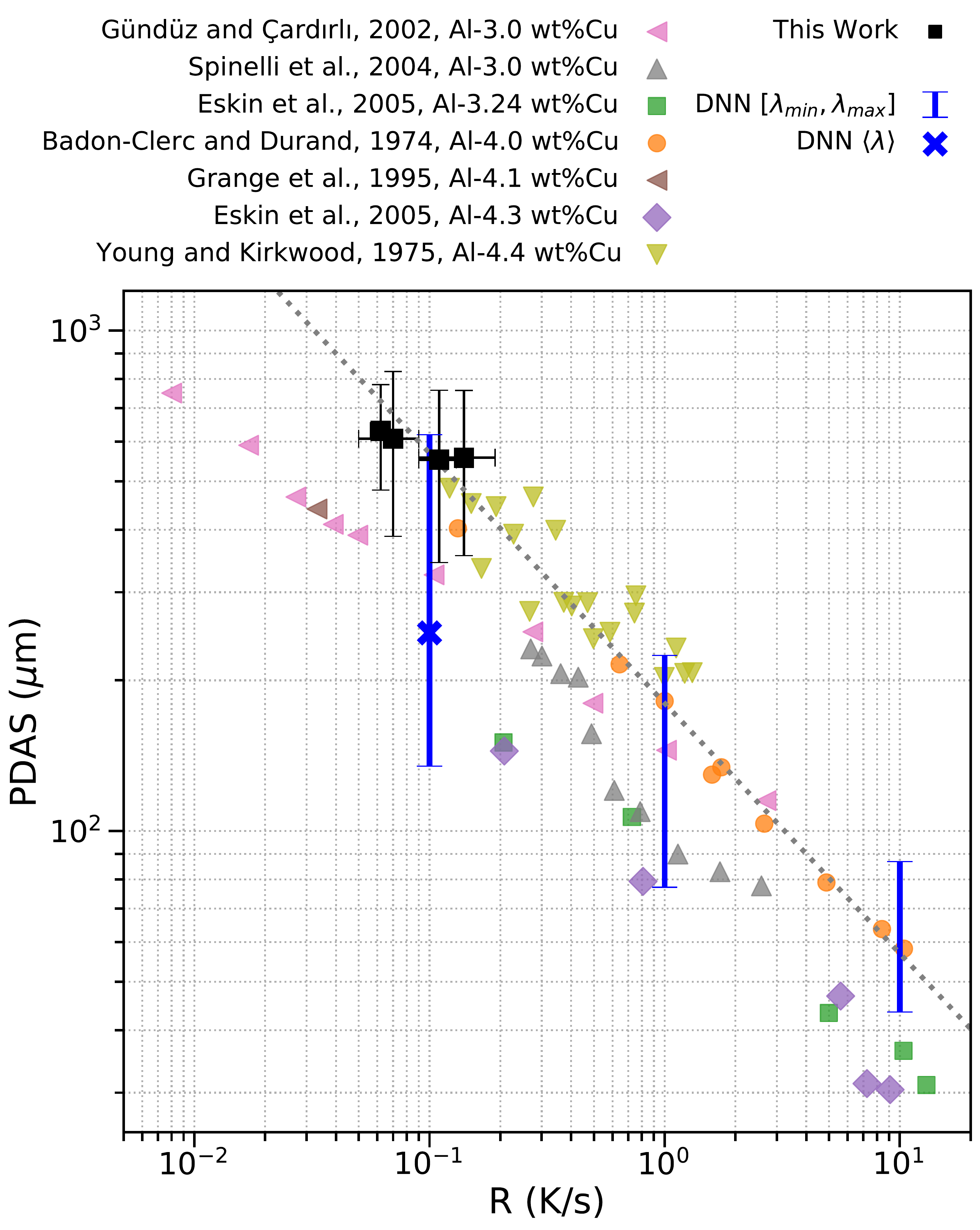}
	\caption{
	Comparison of PDAS as a function of cooling rate for the Al-4\,wt\%Cu alloy in our experiments (both cooling conditions), together with literature results, and DNN (blue) predictions of dendritic stability range (error bars) and average spacing in a spatially extended dendritic array (\thickcross). 
	}
	\label{fig:PDAS_Al4CU}
\end{figure}

\begin{figure}[!t]
	\centering
	\includegraphics[width=7.5cm]{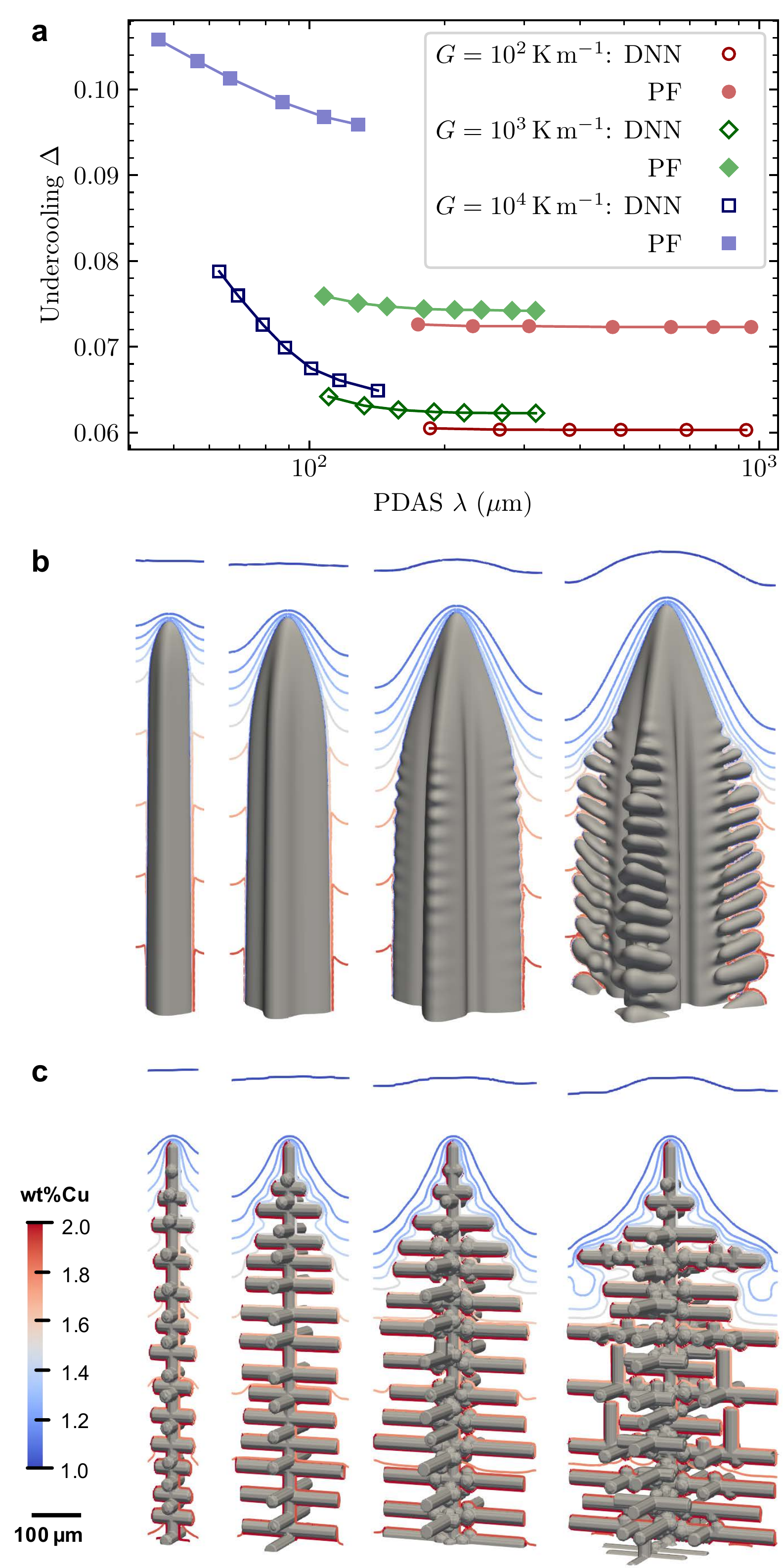}
	\caption{
Simulation results for PDAS stability range for the Al-1\,wt\%Cu alloy at $V=160~\mu$m/s:
(a) Steady-state primary tip undercooling $\Delta$ versus $\lambda$, showing only the stable spacings $\lambda_{min}<\lambda<\lambda_{max}$ predicted by PF (full symbols) and DNN (open symbols) models for $G=10^2$ (red), $10^3$ (green), and $10^4$~K/m (blue), and dendrite morphologies simulated with (b) PF and (c) DNN models for stable spacings $\lambda\approx 110$, 180, 250, and $320~\mu$m (left to right) for $G=10^3~$K/m.
Panels (b) and (c) appear at the same scale, showing the solid-liquid interface and similar iso-concentration lines (wt\%Cu) in a (010) plane.
In panel (a), the leftmost and rightmost data points show the lowest ($\lambda_{min}$) and highest ($\lambda_{max}$) stable spacings identified for each set of parameters, even though within the stability range only a small subset of the simulations are shown as symbols for the sake of readability.
	\label{fig:simu_spac_rang_al1cu}
	}
\end{figure}

\begin{figure}[!t]
	\centering
	\includegraphics[width=7.5cm]{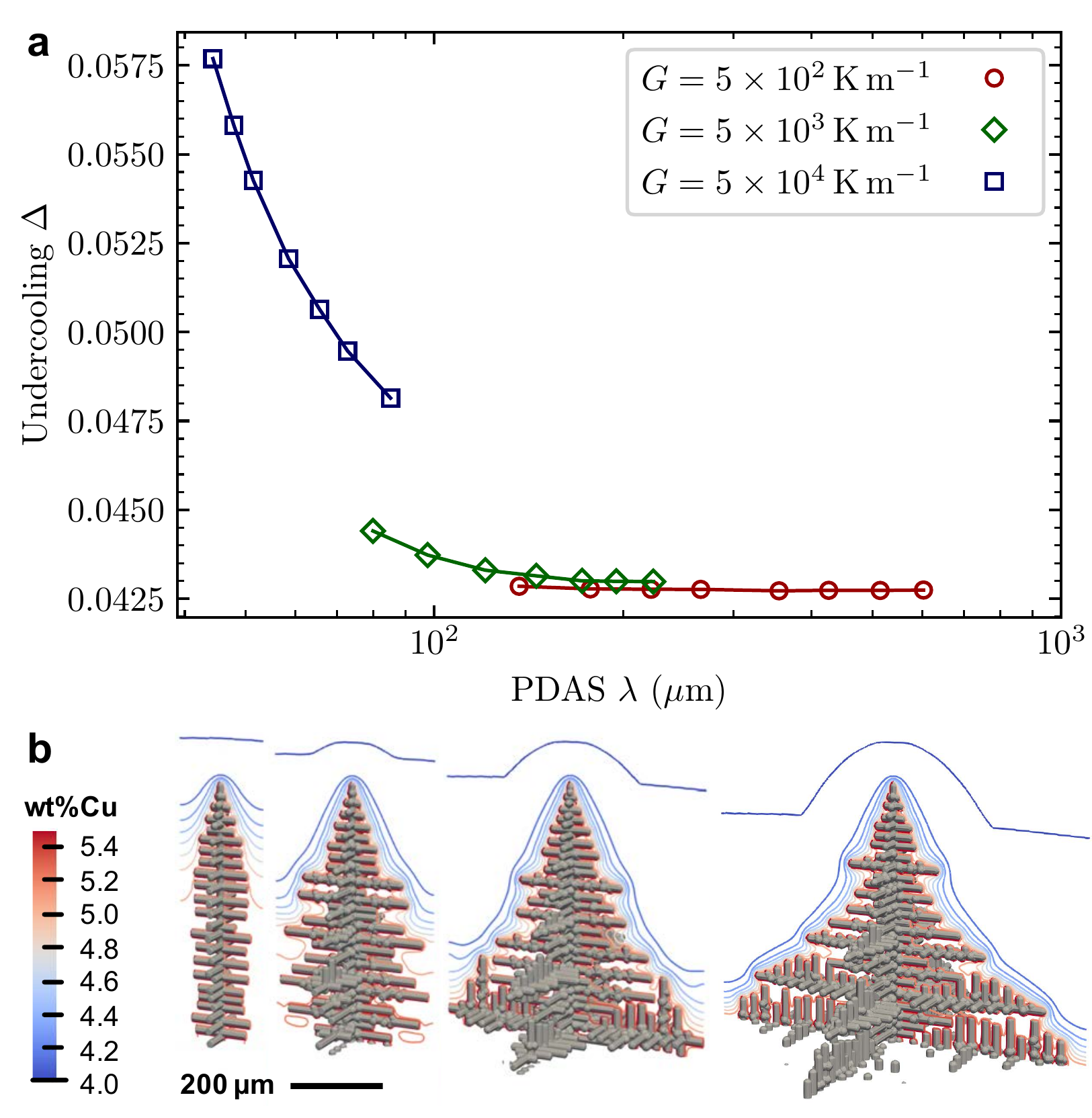}
	\caption{
Simulation results for PDAS stability range for the Al-4\,wt\%Cu alloy at $V=200~\mu$m/s:
(a) Steady-state primary tip undercooling $\Delta$ versus PDAS, showing only the stable spacings $\lambda_{min}<\lambda<\lambda_{max}$ predicted by the DNN simulations for $G=5\times10^2$ (red), $5\times10^3$ (green), and $5\times10^4$~K/m (blue), and (b) view of the dendrite morphologies for stable spacings $\lambda\approx 137$, 296, 454, and 603~$\mu$m (left to right) for $G=5\times10^2~$K/m, showing the solid-liquid interface and iso-concentration lines (wt\%Cu) in a (010) plane.
In panel (a), the leftmost and rightmost data points show the lowest ($\lambda_{min}$) and highest ($\lambda_{max}$) stable spacings identified for each set of parameters, even though within the stability range only a small subset of the simulations are shown as symbols for the sake of readability.
	\label{fig:simu_spac_rang_al4cu}
	}
\end{figure}

For both PF and DNN models, results of single-spacing simulations appear in Figs~\ref{fig:simu_spac_rang_al1cu} (Al-1\,wt\%Cu) and \ref{fig:simu_spac_rang_al4cu} (Al-4\,wt\%Cu), showing (a) primary tip undercooling versus spacing within the stable PDAS range, as well as (b-c) typical primary dendrites within the stable range and their surrounding solute field in a (010) plane. 
We also report the calculated values of $\lambda_{min}$, $\lambda_{max}$, as well as the ratio $\lambda_{max}/\lambda_{min}$ for all simulations in Table~\ref{tab_ratio_max_min_al1cu} (Al-1\,wt\%Cu) and Table~\ref{tab_ratio_max_min_al4cu} (Al-4\,wt\%Cu).

\begin{figure}[!t]
	\centering
	\includegraphics[width=7.5cm]{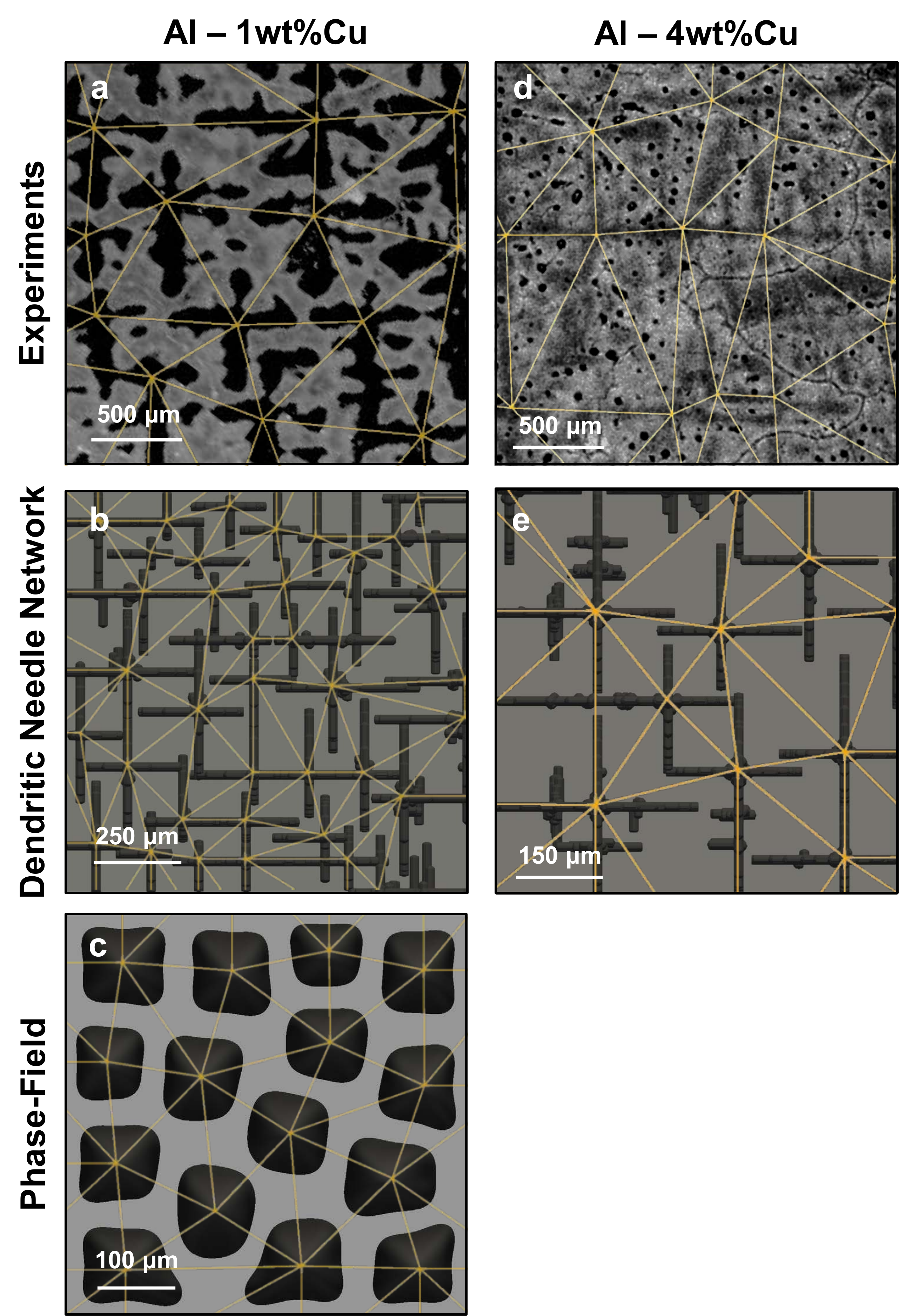}
	\caption{
	Experimental micrographs (a,d) and steady state microstructures observed in the $(x-)$ direction simulated using the DNN (b,e) and PF (c) methods for the Al-1\,wt\%Cu (a-c) and the Al-4\,wt\%Cu (d,e) alloy processed in the presence of an insulating alumina plate between the mold and the chill face, i.e. for $V=160~\mu$m/s and $G=10^3$~K/m for Al-1\,wt\%Cu (a-c) and $V=200~\mu$m/s and $G=5\times10^2$~K/m for Al-4\,wt\%Cu (d,e).	
	\label{fig:simu_spac_array}
	}
\end{figure}

Final microstructures in the spatially extended simulations are illustrated and compared to experimental micrographs in Fig.~\ref{fig:simu_spac_array} for the Al-1\,wt\%Cu (a-c) and the Al-4\,wt\%Cu (d,e) alloys, superimposed with a Delaunay triangulation similar to the one used to calculate average primary spacings.
Average spacings $\langle \lambda \rangle$ from simulations are reported in Figs~\ref{fig:PDAS_Al1CU} and \ref{fig:PDAS_Al4CU} as thick crosses (\thickcross), together with predicted stability ranges as vertical error bars.

\begin{table}[!t]
\centering
\caption{
Predicted spacing stability limits and ratio $\lambda_{max}/\lambda_{min}$ from PF and DNN simulations of Al-1\,wt\%Cu solidification at $V=160~\mu$m/s.
}
\label{tab_ratio_max_min_al1cu}
\begin{tabular}{c c r r r}
\hline
    $G$ (K/m) & & $10^2$ & $10^3$ & $10^4$ \\
\hline
\hline
    $\lambda_{min}$ ($\mu$m) & PF & 161.5 & 107.8 & 46.2 \\
 & DNN & 183.6 & 108.8 & 61.5 \\
    $\lambda_{max}$ ($\mu$m) & PF & 959.0 & 318.1 & 128.3 \\
 & DNN & 940.7 & 320.2 & 143.5 \\
    $\lambda_{max}/\lambda_{min}$ & PF & 5.94 & 2.95 & 2.78 \\
 & DNN & 5.12 & 2.94 & 2.33 \\
\hline
\end{tabular}
\end{table}

\begin{table}[!t]
\centering
\caption{
Predicted spacing stability limits and ratio $\lambda_{max}/\lambda_{min}$ from DNN simulations of the Al-4\,wt\%Cu alloy at $V=200~\mu$m/s.
}
\label{tab_ratio_max_min_al4cu}
\begin{tabular}{c r r r}
\hline
$G$ (K/m) & $5\times10^2$ & $5\times10^3$ & $5\times10^4$ \\
\hline
\hline
    $\lambda_{min}$ ($\mu$m) & 134.8 & 77.2 & 43.5 \\
    $\lambda_{max}$ ($\mu$m) & 619.2 & 224.4 & 86.9 \\
    $\lambda_{max}/\lambda_{min}$  & 4.59 & 2.88 & 1.99 \\
\hline
\end{tabular}
\end{table}

In terms of computational performance (see brief discussion in the Supplementary Material), a representative simulation in the vicinity of $\lambda\approx\lambda_{min}$ at $G=10^3~$K/m for the Al-1\,wt\%Cu alloy, performed on the same hardware (one NVIDIA GeForce$^\circledR$ GTX 1080Ti and a single core of an Intel$^\circledR$ i7-6850K) was performed about six times faster with DNN than PF. 
However, in this regime where both PF and DNN simulations are achievable, DNN simulations are far outside of their optimal performance, which can only be obtained when $\rho\ll D/V$ and PF simulations become unfeasible.

\section{Discussion}
\label{sec_disscussion}

\subsection{Experiments}

The thermocouples farthest from the chill plate, namely TC5, TC6, and TC7 provide an estimation of the vertical thermal gradient. 
The temperatures of TC6 and TC7 mostly overlap, except in the Al-4\,wt\%Cu case with alumina plate (Fig.~\ref{fig:Temp_time}a), where TC7 is at a slightly higher temperature. 
This is likely due to a second solidification front moving from the opposite end of the mold and promoting earlier solidification of that region. 
Something similar occurs in the Al-4\,wt\%Cu alloy when the mold is in direct contact with the chill plate (Fig.~\ref{fig:Temp_time}d): TC5 to TC7 exhibit a faster initial cooling rates than TC4 until the temperatures converge, suggesting that solidification occurs around the same time in this region. 
This uncertainty in the temperature distribution and history makes it more challenging to establish a clear processing-microstructure relationship in this central region of the mold.
Therefore, we chose to focus primarily on the solidification behavior between TC1 and TC2, where the solidification conditions can be estimated with greater confidence.

Dendritic arrays have a more regular arrangement and a better defined microstructure in the micrographs of the central region of the mold (i.e. between TC2 and TC3, or TC3 and TC4) than near the chill plate (i.e. between TC1 and TC2) in the Al-1\,wt\%Cu alloy (see Fig.~\ref{fig:pole100}).  
This is likely a sign of transient thermal conditions at the beginning of solidification.
Moreover, the insulation of the mold is poorer near the chill face, leading to possible heat fluxes in different directions, such that the dendrites growth direction may deviate from the main thermal gradient direction.
Further away from the chill plate, the solidification front is stabilized and these disturbances vanish for the most part. 
In spite of these transient conditions, microstructures between TC1 and TC2 are sufficiently defined to extract a reasonable measure of the PDAS. 

The expected power law dependence of the PDAS with the cooling rate $R$ appears in Figs \ref{fig:PDAS_Al1CU} and \ref{fig:PDAS_Al4CU}, as the data follows the illustrated $R^{-0.5}$ slope suggested in earlier studies  \cite{Badon1974,Okamoto1975,Young1975,Kaneko1977,McCartney1981,Grange1995,Su1998,Lin1999,Quaresma2000,Gunduz2002,Rocha2003,Spinelli2004,Eskin2005}. 
While our experiments did not permit to scan an extensive range of cooling rates, they still suggest a decrease in PDAS for higher cooling rate, consistently with this $\lambda\sim R^{-0.5}$ power law.

For the Al-4\,wt\%Cu alloy, our measured PDAS coincide well with the results of \citep{Badon1974,Su1998,Young1975} for similar cooling rates. 
However, our measurements for both alloys appear on the higher end of the measured spacing distribution from the literature for comparable compositions. 
They also appear on the high end of the calculated stable ranges.
This may be attributed to several factors.

First, the representation of $\lambda$ as a sole function of $R$ is a simplification. 
For a given alloy, the average PDAS is a function of both $G$ and $R$, such that, for a given cooling rate $R=G\times V$, different sets of points might represent substantially different conditions and even different growth regimes. 
A better representation would be a three-dimensional graph with axes $\lambda$, $G$, and $V$ (or even 4D including $c_\infty$). 

Second, experimental setups and configurations compared here differ substantially from one another.
Most previous data comes from thin-sample or Bridgman directional solidification experiments. 
Even among these, whether the solidification is horizontal, vertical upwards, or vertical downwards, has important consequences, particularly in terms of thermo-solutal convection \cite{mathiesen2006crystal, ruvalcaba2007situ, boden2008x, shevchenko2013chimney, bogno2011analysis, clarke2015x, gibbs2016situ}.
For example, in the experiments from Ref.~\cite{clarke2017microstructure} (full orange circles in Fig.~\ref{fig:PDAS_Al1CU}),  performed vertically upwards, it was shown that, in spite of the low sample thickness ($200~\mu$m), the extent of convection was sufficient for these experiments to occur within a cellular-to-dendritic transition regime, explaining the inverse trend with $\partial \lambda/\partial R>0$ in Fig.~\ref{fig:PDAS_Al1CU} \cite{clarke2017microstructure}, in place of the expected $\lambda\sim R^{-0.5}$ in the fully dendritic regime.

Third, we relied on limited local thermocouple measurements, which arguably leaves room for error in the estimation of the local $(G,V)$ conditions at exact times and locations within the mold.
A more accurate estimation of the local temperature profile in the sample could be achieved, for instance extending our multiscale modeling approach to thermal and solutal transport simulations at the macroscopic scale, which is the focus of an ongoing follow-up study.

Finally, we based our estimation of the PDAS on two-dimensional cross-sections.
Even though we attempted to correct for the misorientation of the dendrite growth direction, we expect that underlying assumptions also lead to some amount of error in the PDAS measurements.
Such issues could be tackled in further studies using three-dimensional visualization methods, such as X-ray tomography \cite{limodin2009situ, cai20164d, tourret2017solidification, gibbs2015three, elder2020microstructural}, which would provide an unambiguous representation of dendritic arrays and spacings. 

\subsection{Simulations}

For the Al-1\,wt\%Cu alloy, the calculated values for $\lambda_{min}$ and $\lambda_{max}$ exhibit a striking agreement between PF and DNN methods, in spite of the lower undercooling predicted by DNN compared to PF (Fig.~\ref{fig:simu_spac_rang_al1cu}a), which was already observed and reported in an ongoing benchmark study~\cite{tourret2020mcwasp}. 
Some discrepancy between DNN and PF appears at higher undercooling (i.e. for $G=10^4~$K/m in Fig.~\ref{fig:simu_spac_rang_al1cu}a), which we attribute to the fact that a higher $G$ promotes a transition toward an intermediate regime between cellular and dendritic.
Consequently, the agreement is best at low $G$, when microstructures are well-developed dendrites made of thin hierarchical branches, for which the DNN model was specifically designed. 
For both compositions and both simulation methods, the dependence of both $\lambda_{min}$ and  $\lambda_{max}$ fall close to the phenomenological $R^{-0.5}$ power law (dotted lines in Figs~\ref{fig:PDAS_Al1CU} and \ref{fig:PDAS_Al4CU}).

Hence, the ratio $\lambda_{max}/\lambda_{min}$ (Tables~\ref{tab_ratio_max_min_al1cu} and \ref{tab_ratio_max_min_al4cu}) also presents a striking agreement between the two simulation methods. 
This ratio is as high as 5 or 6 for the lowest values of $G$, and decreases consistently with increasing $G$, as microstructures approach an intermediate cellular-dendritic regime.
The fact that $\lambda_{max}/\lambda_{min}$ is much higher than 2 is consistent with recent studies \cite{tourret2015three, pereda2017experimental, clarke2017microstructure, song2018thermal} and in contrast with earlier theories, which suggested that $\lambda_{max}/\lambda_{min}\approx 2$ \cite{lu1992numerical, hunt1996numerical}, but mostly by lack of a proper simulation method to study branching instability at $\lambda\approx\lambda_{max}$.
The decrease in $\lambda_{max}/\lambda_{min}$ with $G$ and the fact that it may tend toward 2 as the gradient increases are new observations that warrant further investigation. 

Overall, this comparison between simulations and experiments, in conditions relevant to industrial casting, provides a compelling demonstration that physics-based models can be coupled across scales to yield quantitative predictions of key microstructural length scales --- here specifically illustrated for the PDAS. 
However, our simulations, while in good agreement with literature data, seem to underestimate the spacings measured in our experiments.
This is due to error sources in both simulations and experiments, such that some limitations in the current calculations and assumptions deserve to be discussed, as they point toward directions for improvement and further research. 

First, it was previously shown that a microstructural steady state, with a fixed PDAS, may require a substantially longer time than that required to reach a steady undercooling and growth velocity of the solidification front~\cite{clarke2017microstructure}. 
Hence, even though microstructures may appear at steady state, a slow interaction and elimination process might still be occurring, possibly leading to an increase of average spacing at longer time scale.

Then, in the single-spacing simulations, we made the assumption of a quincux (fcc-like) spatial order of the dendritic arrays (Fig.~\ref{sim_sym}).
As such, these simulations are only representative of $\lambda_{min}$ and $\lambda_{max}$ in a perfectly ordered array with this specific arrangement. 
This ordering can be identified in terms of nearest neighbors as an irregular hexagonal pattern (dotted line in Fig.~\ref{sim_sym}), with two longer edges along the directions corresponding to side-branches growth.
It was shown to be predominant in directionally solidified single crystals (see, e.g., the high density of such irregular hexagons in the bottom left region of Fig.~2b in Ref.~\cite{strickland2020nature}) --- in contrast to predominantly regular hexagonal arrays in cellular growth due to the axisymmetry of cells \cite{bergeon2013spatiotemporal, tourret2015oscillatory}.
While we chose to leave this aspect out of the scope of the current study, it would be interesting, among future directions, to investigate the influence of the spatial order on the resulting spacing stability range, as well as the possible transition from regular to irregular hexagonal arrays with the appearance of side-branches between cellular and dendritic growth regimes.

In both experiments and spatially extended simulations, spatial order of the array is never perfect.
In the experiments, array disorder is primarily promoted by the inhomogeneous processing conditions, with gradients in $G$, $V$, and $c_\infty$ due to a large extent to convection, and the fact that microstructures are polycrystalline, thus leading to sustained elimination and branching events \cite{pereda2017experimental, song2018propagative}.

Due to this disorder, a more advanced algorithm than our Delaunay triangulation could have led to a more accurate PDAS estimation~\cite{warnken2011characterization, tschopp2014characterizing, takaki2016primary, strickland2020nature}.
However, we expect that such methods would have only filtered a handful of nearest neighbor outliers, such that the resulting correction would have been minimal once reported within the nearly two orders of magnitude of PDAS variation discussed here (Figs~\ref{fig:PDAS_Al1CU}-\ref{fig:PDAS_Al4CU}).

In all simulations, we made the assumptions that $G$ and $V$ are constant, that heat transport is infinitely fast compared to solute transport, and that solute transport occurs primarily by diffusion in the liquid phase. 
While these are all reasonable assumptions for the solidification of a metallic system considered at a scale lower than the characteristic hydrodynamic length, we expect that such assumptions introduce some amount of discrepancy by deviating from exact experimental conditions.

Finally, one intrinsic limitation of the DNN model is that primary dendrite trunks cannot drift laterally, such that spacings can only be adjusted through elimination and branching events. 
This leads to a distribution of spacing less homogeneous than in both experiments and PF simulations (Fig.~\ref{fig:simu_spac_array}), as already noted for thin-sample solidification of Al-Si alloys~\cite{tourret2015three}. 
The DNN model might be adapted to allow such a drifting, but this would require (1) a new mathematical formulation of the FIF that would incorporate the directionality of the incoming solute flux, (2) establishing laws for dendrite drifting velocities in inhomogeneous arrays, and (3) a notably more complex implementation.

In spite of these limitations, the present comparison between experiments and simulations highlights a promising pathway toward quantitative predictions of microstructural length scales (e.g. PDAS) in casting processes.
The striking agreement between PF and DNN predictions is a strong indication that discrepancies between simulated and experimental data stem from the fundamental assumptions behind the models rather than from the upscaling from PF to DNN simulations.

\section{Summary and perspectives}
\label{sec_conclusion}

In this article, we demonstrated that primary dendritic spacings in metallic alloys cast in industrially relevant conditions can be calculated by directly combining physics-based models --- with barely any adjustable parameters.
On the one hand, we performed casting experiments of Al-Cu alloys and related the cooling conditions measured by thermocouples to the PDAS measured from cross-sections micrographs.
On the other hand, we calculated PDAS stability ranges and simulated entire dendritic arrays using two different methods, namely phase-field (PF) and dendritic needle network (DNN) models.
For the alloy with lower copper content (Al-1\,wt\%Cu), both types of simulations are feasible, and they lead to very similar results.
For the more concentrated alloy (Al-4\,wt\%Cu), the wide separation of scale between dendrite tip radius $\rho$ and diffusion length $l_D$ makes PF simulations prohibitively challenging, while DNN simulations still yield a good agreement with experiments.

We compared our results to the extensive body of literature on PDAS selection in Al-Cu alloys (see Fig.~I in the Supplementary Material), and found an overall good agreement between our results and published data.
It is, to our knowledge, the first time that this broad data set was compiled and directly compared to simulation predictions. 
The agreement between simulations and experimental data provides a conclusive validation of the models' capability to reliably predict primary spacings.
Moreover, in spite of a slight underestimation of undercooling by the DNN method as compared to PF (already identified elsewhere~\cite{tourret2020mcwasp}), the excellent agreement in terms of stable spacing ranges for different sets of parameters provides a further validation of the models. 
In summary, we have shown that quantitative PF simulations, computationally limited to dilute alloys, could be readily extended to more concentrated alloys using the DNN approach. 
This opens the way to further direct comparisons of experiments and simulations at experimentally relevant length and time scales.

Perspectives from this work are manyfold. 
From a fundamental standpoint, the relationship between the width of the PDAS stability range (i.e. the ratio $\lambda_{max}/\lambda_{min}$) and the temperature gradient remains to be explored in further details. 
From an ICME perspective, ongoing works include the extension of this multiscale modeling strategy toward both the macroscopic scale --- e.g. coupling to thermomechanical simulations of the process with possible integration of convection into DNN simulations \cite{tourret2019multiscale,isensee2020three} --- and toward the nanoscale --- e.g. calculating phase diagrams from first principles \cite{liu2020prediction, liu2020mgzn} and solid-liquid interface properties from molecular dynamics \cite{hoyt2003atomistic, brown2017interfacial}.
Such computational frameworks could also be complemented by similar multi-scale strategies for later processing steps, e.g. including the formation of secondary phases such as precipitates \cite{liu2017multiscale, liu2019precipitation}.

\section*{Acknowledgements}
This research was funded by the European Research Council Advanced Grant VIRMETAL under the European Union's Horizon 2020 research and innovation programme (Grant Agreement 669141).
D.T. acknowledges support from the European Union's Horizon 2020 research and innovation programme through a Marie Sk\l odowska-Curie Individual Fellowship (Grant Agreement 842795). Additional support from  the HexaGB project of the Spanish Ministry of Science (reference RTI2018-098245) is also gratefully acknowledged. B.B. acknowledges support by the Spanish Ministry of Education through the Fellowship FPU15/00403 and D.T. and J.LL. also gratefully acknowledge support by the NVIDIA Corporation with the donation of two Titan Xp GPUs. 
O.M.W. and M.S.T. would like to acknowledge Purdue University for supporting this work through start-up funds.
The valuable help from undergraduate students Jes\'us Mart\'inez Justo and Luis Montes Mota (Polytechnic University of Madrid) for the metallography sample preparation is also gratefully acknowledged.



\begin{thebibliography}{100}
\expandafter\ifx\csname url\endcsname\relax
  \def\url#1{\texttt{#1}}\fi
\expandafter\ifx\csname urlprefix\endcsname\relax\def\urlprefix{URL }\fi
\expandafter\ifx\csname href\endcsname\relax
  \def\href#1#2{#2} \def\path#1{#1}\fi

\bibitem{flemings1974solidification}
M.~C. Flemings, Solidification processing, Metallurgical transactions 5~(10)
  (1974) 2121--2134.

\bibitem{Kurz2019}
W.~Kurz, D.~J. Fisher, R.~Trivedi, {Progress in modelling solidification
  microstructures in metals and alloys: dendrites and cells from 1700 to 2000},
  International Materials Reviews 64~(6) (2019) 311--354.

\bibitem{Trivedi1994}
R.~Trivedi, W.~Kurz, {Dendritic growth}, International Materials Reviews 39~(2)
  (1994) 49--74.

\bibitem{quaresma2000correlation}
J.~M. Quaresma, C.~A. Santos, A.~Garcia, Correlation between unsteady-state
  solidification conditions, dendrite spacings, and mechanical properties of
  al-cu alloys, Metallurgical and Materials Transactions A 31~(12) (2000)
  3167--3178.

\bibitem{osorio2002modeling}
W.~R. Os{\'o}rio, A.~Garcia, Modeling dendritic structure and mechanical
  properties of zn--al alloys as a function of solidification conditions,
  Materials Science and Engineering: A 325~(1-2) (2002) 103--111.

\bibitem{osorio2006effect}
W.~R. Osorio, P.~R. Goulart, A.~Garcia, G.~A. Santos, C.~M. Neto, Effect of
  dendritic arm spacing on mechanical properties and corrosion resistance of al
  9 wt pct si and zn 27 wt pct al alloys, Metallurgical and Materials
  Transactions A 37~(8) (2006) 2525--2538.

\bibitem{beckermann2002modelling}
C.~Beckermann, Modelling of macrosegregation: applications and future needs,
  International Materials Reviews 47~(5) (2002) 243--261.

\bibitem{osorio2005effect}
W.~R. Os{\'o}rio, C.~M. Freire, A.~Garcia, The effect of the dendritic
  microstructure on the corrosion resistance of zn--al alloys, Journal of
  Alloys and Compounds 397~(1-2) (2005) 179--191.

\bibitem{osorio2007roles}
W.~Os{\'o}rio, J.~Spinelli, I.~Ferreira, A.~Garcia, The roles of
  macrosegregation and of dendritic array spacings on the electrochemical
  behavior of an al--4.5 wt.\% cu alloy, Electrochimica Acta 52~(9) (2007)
  3265--3273.

\bibitem{nasser1985flow}
R.~Nasser-Rafi, R.~Deshmukh, D.~R. Poirier, Flow of interdendritic liquid and
  permeability in pb-20 wt pct sn alloys, Metallurgical Transactions A 16~(12)
  (1985) 2263--2271.

\bibitem{poirier1987permeability}
D.~R. Poirier, Permeability for flow of interdendritic liquid in
  columnar-dendritic alloys, Metallurgical Transactions B 18~(1) (1987)
  245--255.

\bibitem{ganesan1992permeability}
S.~Ganesan, C.~Chan, D.~R. Poirier, Permeability for flow parallel to primary
  dendrite arms, Materials Science and Engineering: A 151~(1) (1992) 97--105.

\bibitem{santos2005permeability}
R.~Santos, M.~Melo, Permeability of interdendritic channels, Materials Science
  and Engineering: A 391~(1-2) (2005) 151--158.

\bibitem{takaki2019permeability}
T.~Takaki, S.~Sakane, M.~Ohno, Y.~Shibuta, T.~Aoki, Permeability prediction for
  flow normal to columnar solidification structures by large--scale simulations
  of phase--field and lattice boltzmann methods, Acta Materialia 164 (2019)
  237--249.

\bibitem{langer1980instabilities}
J.~S. Langer, Instabilities and pattern formation in crystal growth, Reviews of
  modern physics 52~(1) (1980) 1.

\bibitem{ben1984pattern}
E.~Ben-Jacob, N.~Goldenfeld, B.~Kotliar, J.~Langer, Pattern selection in
  dendritic solidification, Physical review letters 53~(22) (1984) 2110.

\bibitem{benamar1993theory}
M.~Ben~Amar, E.~Brener, Theory of pattern selection in three-dimensional
  nonaxisymmetric dendritic growth, Physical review letters 71~(4) (1993) 589.

\bibitem{brener1993needle}
E.~Brener, Needle-crystal solution in three-dimensional dendritic growth,
  Physical review letters 71~(22) (1993) 3653.

\bibitem{karma2000three}
A.~Karma, Y.~H. Lee, M.~Plapp, Three-dimensional dendrite-tip morphology at low
  undercooling, Physical Review E 61~(4) (2000) 3996.

\bibitem{somboonsuk1985dynamical}
K.~Somboonsuk, R.~Trivedi, Dynamical studies of dendritic growth, Acta
  Metallurgica 33~(6) (1985) 1051--1060.

\bibitem{trivedi1985pattern}
R.~Trivedi, K.~Somboonsuk, Pattern formation during the directional
  solidification of binary systems, Acta metallurgica 33~(6) (1985) 1061--1068.

\bibitem{Kurz1981}
W.~Kurz, D.~J. Fisher, {Dendrite growth at the limit of stability: tip radius
  and spacing}, Acta Metallurgica 29~(1) (1981) 11--20.

\bibitem{lu1992numerical}
S.-Z. Lu, J.~Hunt, A numerical analysis of dendritic and cellular array growth:
  the spacing adjustment mechanisms, Journal of Crystal Growth 123~(1-2) (1992)
  17--34.

\bibitem{hunt1996numerical}
J.~Hunt, S.-Z. Lu, Numerical modeling of cellular/dendritic array growth:
  spacing and structure predictions, Metallurgical and Materials Transactions A
  27~(3) (1996) 611--623.

\bibitem{warren1993prediction}
J.~A. Warren, J.~Langer, Prediction of dendritic spacings in a
  directional-solidification experiment, Physical Review E 47~(4) (1993) 2702.

\bibitem{HT1994}
S.~H. Han, R.~Trivedi, {Primary spacing selection in directionally solidified
  alloys}, Acta Metallurgica Et Materialia 42~(1) (1994) 25--41.

\bibitem{weidong1993primary}
H.~Weidong, G.~Xingguo, Z.~Yaohe, Primary spacing selection of constrained
  dendritic growth, Journal of crystal growth 134~(1-2) (1993) 105--115.

\bibitem{Trivedi1994a}
R.~Trivedi, W.~Kurz, {Solidification microstructures: A conceptual approach},
  Acta Metallurgica et Materialia 42~(1) (1994) 15--23.

\bibitem{KurzFisher}
W.~Kurz, D.~J. Fisher, Fundamentals of solidification, Trans Tech Publ, 1989.

\bibitem{Hunt1979}
J.~Hunt, Solidification and casting of metals, The Metal Society, London 3.

\bibitem{Boettinger2000}
W.~J. Boettinger, S.~R. Coriell, A.~L. Greer, A.~Karma, W.~Kurz, M.~Rappaz,
  R.~Trivedi, {Solidification microstructures: recent developments, future
  directions}, Acta Materialia 48 (2000) 43--70.

\bibitem{Provatas2010}
N.~Provatas, K.~Elder, {Phase-Field Methods in Materials Science and
  Engineering}, Wiley-VCH Verlag GmbH {\&} Co., 2010.

\bibitem{tourret2015oscillatory}
D.~Tourret, J.-M. Debierre, Y.~Song, F.~L. Mota, N.~Bergeon, R.~Guerin,
  R.~Trivedi, B.~Billia, A.~Karma, Oscillatory cellular patterns in
  three-dimensional directional solidification, Physical Review E 92~(4) (2015)
  042401.

\bibitem{tourret2015growth}
D.~Tourret, A.~Karma, {Growth competition of columnar dendritic grains: A
  phase-field study}, Acta Materialia 82 (2015) 64--83.

\bibitem{Geying1987}
A.~Geying, L.~Lixin, {Dendrite Spacing in Unidirectionally Solidified Al-Cu
  Alloy}, Journal of Crystal Growth 80 (1987) 383--392.

\bibitem{McCartney1981}
D.~G. McCartney, J.~D. Hunt, {Measurements of Cell and Primary Dendrite}, Acta
  Metallurgica 29~(11) (1981) 1851--1863.

\bibitem{Lin1999}
X.~Lin, W.~Huang, J.~Feng, T.~Li, Y.~Zhou, {History-dependent selection of
  primary cellular/dendritic spacing during unidirectional solidification in
  aluminum alloys}, Acta Materialia 47~(11) (1999) 3271--3280.

\bibitem{Gunduz2002}
M.~G{\"{u}}nd{\"{u}}z, E.~{\c{C}}adırlı, {Directional solidification of
  aluminium–copper alloys}, Materials Science and Engineering: A 327~(2)
  (2002) 167--185.

\bibitem{Okamoto1975}
T.~Okamoto, K.~Kishitake, Dendritic structure in unidirectionally solidified
  aluminum, tin, and zinc base binary alloys, Journal of Crystal Growth 29~(2)
  (1975) 137 -- 146.

\bibitem{Young1975}
K.~P. Young, D.~H. Kerkwood, {The dendrite arm spacings of aluminum-copper
  alloys solidified under steady-state conditions}, Metallurgical Transactions
  A 6 (1975) 197--205.

\bibitem{Kaneko1977}
J.~Kaneko, {Dendrite coarsening during solidification of hypo-and
  hyper-eutectic AI-Cu alloys}, Journal of Materials Science 12 (1977)
  1392--1400.

\bibitem{Su1998}
R.-J. Su, R.~A. Overfelt, W.~A. Jemian, {Microstructural and compositional
  transients during accelerated directional solidification of Al-4.5 wt pct
  Cu}, Metallurgical and Materials Transactions A 29 (1998) 2375--2381.

\bibitem{Grange1995}
G.~Grange, J.~Gastaldi, C.~Jourdan, B.~Billia, Evolution of characteristic
  pattern parameters in directional solidification of thin samples of a dilute
  alcu alloy, Journal of crystal growth 151~(1-2) (1995) 192--199.

\bibitem{Quaresma2000}
J.~M.~V. Quaresma, C.~A. Santos, A.~Garcia, {Correlation between unsteady-state
  solidification conditions, dendrite spacings, and mechanical properties of
  Al-Cu alloys}, Metallurgical and Materials Transactions A 31~(12) (2000)
  3167--3178.

\bibitem{Rocha2003}
O.~L. Rocha, C.~A. Siqueira, A.~Garcia, {Heat flow parameters affecting
  dendrite spacings during unsteady-state solidification of Sn-Pb and Al-Cu
  alloys}, Metallurgical and Materials Transactions A 34~(4) (2003) 995--1006.

\bibitem{Spinelli2004}
J.~E. Spinelli, D.~M. Rosa, I.~L. Ferreira, A.~Garcia, {Influence of melt
  convection on dendritic spacings of downward unsteady-state directionally
  solidified Al–Cu alloys}, Materials Science and Engineering: A 383~(2)
  (2004) 271--282.

\bibitem{Eskin2005}
D.~Eskin, Q.~Du, D.~Ruvalcaba, L.~Katgerman, {Experimental study of structure
  formation in binary Al-Cu alloys at different cooling rates}, Materials
  Science and Engineering A 405~(1-2) (2005) 1--10.

\bibitem{rappaz2010numerical}
M.~Rappaz, M.~Bellet, M.~Deville, Numerical modeling in materials science and
  engineering, Vol.~32, Springer Science \& Business Media, 2010.

\bibitem{ni1991volume}
J.~Ni, C.~Beckermann, A volume-averaged two-phase model for transport phenomena
  during solidification, Metallurgical Transactions B 22~(3) (1991) 349--361.

\bibitem{wang1996equiaxed_1}
C.~Wang, C.~Beckermann, Equiaxed dendritic solidification with convection: Part
  i. multiscale/multiphase modeling, Metallurgical and materials transactions A
  27~(9) (1996) 2754--2764.

\bibitem{wang1996equiaxed_2}
C.~Wang, C.~Beckermann, Equiaxed dendritic solidification with convection: Part
  ii. numerical simulations for an al-4 wt pct cu alloy, Metallurgical and
  Materials Transactions A 27~(9) (1996) 2765--2783.

\bibitem{combeau2009prediction}
H.~Combeau, M.~Zalo{\v{z}}nik, S.~Hans, P.~E. Richy, Prediction of
  macrosegregation in steel ingots: Influence of the motion and the morphology
  of equiaxed grains, Metallurgical and materials transactions B 40~(3) (2009)
  289--304.

\bibitem{wu2009modeling_1}
M.~Wu, A.~Ludwig, Modeling equiaxed solidification with melt convection and
  grain sedimentation i: Model description, Acta Materialia 57~(19) (2009)
  5621--5631.

\bibitem{wu2009modeling_2}
M.~Wu, A.~Ludwig, Modeling equiaxed solidification with melt convection and
  grain sedimentation ii. model verification, Acta Materialia 57~(19) (2009)
  5632--5644.

\bibitem{tourret2011multiple}
D.~Tourret, C.-A. Gandin, T.~Volkmann, D.~M. Herlach, Multiple non-equilibrium
  phase transformations: Modeling versus electro-magnetic levitation
  experiment, Acta materialia 59~(11) (2011) 4665--4677.

\bibitem{tourret2011gas}
D.~Tourret, G.~Reinhart, C.-A. Gandin, G.~Iles, U.~Dahlborg, M.~Calvo-Dahlborg,
  C.~Bao, Gas atomization of al--ni powders: Solidification modeling and
  neutron diffraction analysis, Acta Materialia 59~(17) (2011) 6658--6669.

\bibitem{rappaz1993probabilistic}
M.~Rappaz, C.-A. Gandin, Probabilistic modelling of microstructure formation in
  solidification processes, Acta metallurgica et materialia 41~(2) (1993)
  345--360.

\bibitem{gandin1994coupled}
C.-A. Gandin, M.~Rappaz, A coupled finite element-cellular automaton model for
  the prediction of dendritic grain structures in solidification processes,
  Acta metallurgica et materialia 42~(7) (1994) 2233--2246.

\bibitem{wang2003model}
W.~Wang, P.~D. Lee, M.~Mclean, A model of solidification microstructures in
  nickel-based superalloys: predicting primary dendrite spacing selection, Acta
  materialia 51~(10) (2003) 2971--2987.

\bibitem{steinbach1999three}
I.~Steinbach, C.~Beckermann, B.~Kauerauf, Q.~Li, J.~Guo, Three-dimensional
  modeling of equiaxed dendritic growth on a mesoscopic scale, Acta Materialia
  47~(3) (1999) 971--982.

\bibitem{viardin2017mesoscopic}
A.~Viardin, M.~Zalo{\v{z}}nik, Y.~Souhar, M.~Apel, H.~Combeau, Mesoscopic
  modeling of spacing and grain selection in columnar dendritic solidification:
  Envelope versus phase-field model, Acta Materialia 122 (2017) 386--399.

\bibitem{tourret2013multiscale}
D.~Tourret, A.~Karma, Multiscale dendritic needle network model of alloy
  solidification, Acta materialia 61~(17) (2013) 6474--6491.

\bibitem{tourret2016three}
D.~Tourret, A.~Karma, Three-dimensional dendritic needle network model for
  alloy solidification, Acta Materialia 120 (2016) 240--254.

\bibitem{tourret2015three_1}
D.~Tourret, A.~J. Clarke, S.~D. Imhoff, P.~J. Gibbs, J.~W. Gibbs, A.~Karma,
  Three-dimensional multiscale modeling of dendritic spacing selection during
  al-si directional solidification, JOM 67~(8) (2015) 1776--1785.

\bibitem{tourret2015three_2}
D.~Tourret, A.~Karma, A.~J. Clarke, P.~J. Gibbs, S.~D. Imhoff,
  Three-dimensional dendritic needle network model with application to al-cu
  directional solidification experiments, {IOP} Conference Series: Materials
  Science and Engineering 84 (2015) 012082.

\bibitem{echebarria2010onset}
B.~Echebarria, A.~Karma, S.~Gurevich, Onset of sidebranching in directional
  solidification, Physical Review E 81~(2) (2010) 021608.

\bibitem{gurevich2010phase}
S.~Gurevich, A.~Karma, M.~Plapp, R.~Trivedi, Phase-field study of
  three-dimensional steady-state growth shapes in directional solidification,
  Physical Review E 81~(1) (2010) 011603.

\bibitem{clarke2017microstructure}
A.~Clarke, D.~Tourret, Y.~Song, S.~Imhoff, P.~Gibbs, J.~Gibbs, K.~Fezzaa,
  A.~Karma, Microstructure selection in thin-sample directional solidification
  of an al-cu alloy: In situ x-ray imaging and phase-field simulations, Acta
  Materialia 129 (2017) 203--216.

\bibitem{boettinger2002phase}
W.~J. Boettinger, J.~A. Warren, C.~Beckermann, A.~Karma, Phase-field simulation
  of solidification, Annual Review of Materials Research 32~(1) (2002)
  163--194.

\bibitem{steinbach2009phase}
I.~Steinbach, Phase-field models in materials science, Modelling and simulation
  in materials science and engineering 17~(7) (2009) 073001.

\bibitem{karma2016atomistic}
A.~Karma, D.~Tourret, Atomistic to continuum modeling of solidification
  microstructures, Current Opinion in Solid State and Materials Science 20~(1)
  (2016) 25--36.

\bibitem{ode2001numerical}
M.~Ode, S.~G. Kim, W.~T. Kim, T.~Suzuki, Numerical prediction of the secondary
  dendrite arm spacing using a phase-field model, ISIJ international 41~(4)
  (2001) 345--349.

\bibitem{karma1996phase}
A.~Karma, W.-J. Rappel, Phase-field method for computationally efficient
  modeling of solidification with arbitrary interface kinetics, Physical review
  E 53~(4) (1996) R3017.

\bibitem{karma1998quantitative}
A.~Karma, W.-J. Rappel, Quantitative phase-field modeling of dendritic growth
  in two and three dimensions, Physical review E 57~(4) (1998) 4323.

\bibitem{karma2001phase}
A.~Karma, Phase-field formulation for quantitative modeling of alloy
  solidification, Physical Review Letters 87~(11) (2001) 115701.

\bibitem{echebarria2004quantitative}
B.~Echebarria, R.~Folch, A.~Karma, M.~Plapp, Quantitative phase-field model of
  alloy solidification, Physical Review E 70~(6) (2004) 061604.

\bibitem{shimokawabe2011peta}
T.~Shimokawabe, T.~Aoki, T.~Takaki, T.~Endo, A.~Yamanaka, N.~Maruyama,
  A.~Nukada, S.~Matsuoka, Peta-scale phase-field simulation for dendritic
  solidification on the tsubame 2.0 supercomputer, in: Proceedings of 2011
  International Conference for High Performance Computing, Networking, Storage
  and Analysis, 2011, pp. 1--11.

\bibitem{shibuta2015solidification}
Y.~Shibuta, M.~Ohno, T.~Takaki, Solidification in a supercomputer: from crystal
  nuclei to dendrite assemblages, Jom 67~(8) (2015) 1793--1804.

\bibitem{provatas1998efficient}
N.~Provatas, N.~Goldenfeld, J.~Dantzig, Efficient computation of dendritic
  microstructures using adaptive mesh refinement, Physical Review Letters
  80~(15) (1998) 3308.

\bibitem{provatas1999adaptive}
N.~Provatas, N.~Goldenfeld, J.~Dantzig, Adaptive mesh refinement computation of
  solidification microstructures using dynamic data structures, Journal of
  computational physics 148~(1) (1999) 265--290.

\bibitem{greenwood2018quantitative}
M.~Greenwood, K.~Shampur, N.~Ofori-Opoku, T.~Pinomaa, L.~Wang, S.~Gurevich,
  N.~Provatas, Quantitative 3d phase field modelling of solidification using
  next-generation adaptive mesh refinement, Computational Materials Science 142
  (2018) 153--171.

\bibitem{rosam2008adaptive}
J.~Rosam, P.~Jimack, A.~Mullis, An adaptive, fully implicit multigrid
  phase-field model for the quantitative simulation of non-isothermal binary
  alloy solidification, Acta Materialia 56~(17) (2008) 4559--4569.

\bibitem{bollada2015three}
P.~Bollada, C.~E. Goodyer, P.~K. Jimack, A.~M. Mullis, F.~Yang, Three
  dimensional thermal-solute phase field simulation of binary alloy
  solidification, Journal of Computational Physics 287 (2015) 130--150.

\bibitem{tourret2019multiscale}
D.~Tourret, M.~M. Francois, A.~J. Clarke, Multiscale dendritic needle network
  model of alloy solidification with fluid flow, Computational Materials
  Science 162 (2019) 206--227.

\bibitem{isensee2020three}
T.~Isensee, D.~Tourret, Three-dimensional needle network model for dendritic
  growth with fluid flow, {IOP} Conference Series: Materials Science and
  Engineering 861 (2020) 012049.

\bibitem{tourret2020comparing}
D.~Tourret, L.~Sturz, A.~Viardin, M.~Zalo{\v{z}}nik, Comparing mesoscopic
  models for dendritic growth, {IOP} Conference Series: Materials Science and
  Engineering 861 (2020) 012002.

\bibitem{tourret2015three_iop}
D.~Tourret, A.~Karma, A.~Clarke, P.~Gibbs, S.~Imhoff, Three-dimensional
  dendritic needle network model with application to al-cu directional
  solidification experiments, in: IOP Conference Series: Materials Science and
  Engineering, Vol.~84, IOP Publishing, 2015, p. 012082.

\bibitem{tourret2015three_jom}
D.~Tourret, A.~J. Clarke, S.~D. Imhoff, P.~J. Gibbs, J.~W. Gibbs, A.~Karma,
  Three-dimensional multiscale modeling of dendritic spacing selection during
  al-si directional solidification, JOM 67~(8) (2015) 1776--1785.

\bibitem{hielscher2019denoising}
R.~Hielscher, C.~B. Silbermann, E.~Schmidl, J.~Ihlemann, Denoising of crystal
  orientation maps, Journal of Applied Crystallography 52~(5).

\bibitem{pereda2017experimental}
J.~Pereda, F.~Mota, L.~Chen, B.~Billia, D.~Tourret, Y.~Song, J.-M. Debierre,
  R.~Gu{\'e}rin, A.~Karma, R.~Trivedi, et~al., Experimental observation of
  oscillatory cellular patterns in three-dimensional directional
  solidification, Physical Review E 95~(1) (2017) 012803.

\bibitem{song2018thermal}
Y.~Song, D.~Tourret, F.~Mota, J.~Pereda, B.~Billia, N.~Bergeon, R.~Trivedi,
  A.~Karma, Thermal-field effects on interface dynamics and microstructure
  selection during alloy directional solidification, Acta Materialia 150 (2018)
  139--152.

\bibitem{akamatsu1997similarity}
S.~Akamatsu, T.~Ihle, Similarity law for the tilt angle of dendrites in
  directional solidification of non-axially-oriented crystals, Physical Review
  E 56~(4) (1997) 4479.

\bibitem{deschamps2008growth}
J.~Deschamps, M.~Georgelin, A.~Pocheau, Growth directions of microstructures in
  directional solidification of crystalline materials, Physical Review E 78~(1)
  (2008) 011605.

\bibitem{barbieri1989predictions}
A.~Barbieri, J.~Langer, Predictions of dendritic growth rates in the linearized
  solvability theory, Physical Review A 39~(10) (1989) 5314.

\bibitem{Badon1974}
M.~Badon-Clerc, F.~Durand, The influence of concentration and thermal
  conditions on the structure of basaltic dendrites in aluminum-copper alloys,
  Mem. Sci. Rev. Metall. 71~(7) (1974) 451--459.

\bibitem{mathiesen2006crystal}
R.~Mathiesen, L.~Arnberg, P.~Bleuet, A.~Somogyi, Crystal fragmentation and
  columnar-to-equiaxed transitions in {Al-Cu} studied by synchrotron x-ray
  video microscopy, Metallurgical and Materials Transactions A 37~(8) (2006)
  2515--2524.

\bibitem{ruvalcaba2007situ}
D.~Ruvalcaba, R.~Mathiesen, D.~Eskin, L.~Arnberg, L.~Katgerman, In situ
  observations of dendritic fragmentation due to local solute-enrichment during
  directional solidification of an aluminum alloy, Acta Materialia 55~(13)
  (2007) 4287--4292.

\bibitem{boden2008x}
S.~Boden, S.~Eckert, B.~Willers, G.~Gerbeth, X-ray radioscopic visualization of
  the solutal convection during solidification of a ga-30 wt pct in alloy,
  Metallurgical and Materials Transactions A 39~(3) (2008) 613--623.

\bibitem{shevchenko2013chimney}
N.~Shevchenko, S.~Boden, G.~Gerbeth, S.~Eckert, Chimney formation in
  solidifying ga-25wt pct in alloys under the influence of thermosolutal melt
  convection, Metallurgical and Materials Transactions A 44~(8) (2013)
  3797--3808.

\bibitem{bogno2011analysis}
A.~Bogno, H.~Nguyen-Thi, A.~Buffet, G.~Reinhart, B.~Billia,
  N.~Mangelinck-No{\"e}l, N.~Bergeon, J.~Baruchel, T.~Schenk, Analysis by
  synchrotron x-ray radiography of convection effects on the dynamic evolution
  of the solid--liquid interface and on solute distribution during the initial
  transient of solidification, Acta Materialia 59~(11) (2011) 4356--4365.

\bibitem{clarke2015x}
A.~J. Clarke, D.~Tourret, S.~D. Imhoff, P.~J. Gibbs, K.~Fezzaa, J.~C. Cooley,
  W.-K. Lee, A.~Deriy, B.~M. Patterson, P.~A. Papin, et~al., X-ray imaging and
  controlled solidification of {Al-Cu} alloys toward microstructures by design,
  Advanced Engineering Materials 17~(4) (2015) 454--459.

\bibitem{gibbs2016situ}
J.~W. Gibbs, D.~Tourret, P.~J. Gibbs, S.~D. Imhoff, M.~J. Gibbs, B.~A. Walker,
  K.~Fezzaa, A.~J. Clarke, In situ x-ray observations of dendritic
  fragmentation during directional solidification of a sn-bi alloy, Jom 68~(1)
  (2016) 170--177.

\bibitem{limodin2009situ}
N.~Limodin, L.~Salvo, E.~Boller, M.~Su{\'e}ry, M.~Felberbaum,
  S.~Gailli{\`e}gue, K.~Madi, In situ and real-time 3-d microtomography
  investigation of dendritic solidification in an al--10 wt.\% cu alloy, Acta
  Materialia 57~(7) (2009) 2300--2310.

\bibitem{cai20164d}
B.~Cai, J.~Wang, A.~Kao, K.~Pericleous, A.~Phillion, R.~Atwood, P.~Lee, 4d
  synchrotron x-ray tomographic quantification of the transition from cellular
  to dendrite growth during directional solidification, Acta Materialia 117
  (2016) 160--169.

\bibitem{tourret2017solidification}
D.~Tourret, J.~Mertens, E.~Lieberman, S.~Imhoff, J.~Gibbs, K.~Henderson,
  K.~Fezzaa, A.~Deriy, T.~Sun, R.~Lebensohn, et~al., From solidification
  processing to microstructure to mechanical properties: A multi-scale x-ray
  study of an al-cu alloy sample, Metallurgical and Materials Transactions A
  48~(11) (2017) 5529--5546.

\bibitem{gibbs2015three}
J.~Gibbs, K.~A. Mohan, E.~Gulsoy, A.~Shahani, X.~Xiao, C.~Bouman, M.~De~Graef,
  P.~Voorhees, The three-dimensional morphology of growing dendrites,
  Scientific reports 5~(1) (2015) 1--9.

\bibitem{elder2020microstructural}
K.~L. Elder, T.~Stan, Y.~Sun, X.~Xiao, P.~W. Voorhees, Microstructural
  characterization of dendritic evolution using two-point statistics, Scripta
  Materialia 182 (2020) 81--85.

\bibitem{tourret2020mcwasp}
D.~Tourret, T.~Isensee, L.~Sturz, A.~Viardin, M.~Zalo{\v{z}}nik, Comparing
  mesoscopic models for dendritic growth, conference MCWASP XV, Stockholm,
  Sweden (virtual conference), June 2020 (2020).

\bibitem{tourret2015three}
D.~Tourret, A.~J. Clarke, S.~D. Imhoff, P.~J. Gibbs, J.~W. Gibbs, A.~Karma,
  Three-dimensional multiscale modeling of dendritic spacing selection during
  {Al-Si} directional solidification, JOM 67~(8) (2015) 1776--1785.

\bibitem{strickland2020nature}
J.~Strickland, B.~Nenchev, S.~Perry, K.~Tassenberg, S.~Gill, C.~Panwisawas,
  H.~Dong, N.~D'Souza, S.~Irwin, On the nature of hexagonality within the
  solidification structure of single crystal alloys: Mechanisms and
  applications, Acta Materialia 200 (2020) 417--431.

\bibitem{bergeon2013spatiotemporal}
N.~Bergeon, D.~Tourret, L.~Chen, J.-M. Debierre, R.~Gu{\'e}rin, A.~Ramirez,
  B.~Billia, A.~Karma, R.~Trivedi, Spatiotemporal dynamics of oscillatory
  cellular patterns in three-dimensional directional solidification, Physical
  review letters 110~(22) (2013) 226102.

\bibitem{song2018propagative}
Y.~Song, S.~Akamatsu, S.~Bottin-Rousseau, A.~Karma, Propagative selection of
  tilted array patterns in directional solidification, Physical Review
  Materials 2~(5) (2018) 053403.

\bibitem{warnken2011characterization}
N.~Warnken, R.~C. Reed, On the characterization of directionally solidified
  dendritic microstructures, Metallurgical and Materials Transactions A 42~(6)
  (2011) 1675--1683.

\bibitem{tschopp2014characterizing}
M.~A. Tschopp, J.~D. Miller, A.~L. Oppedal, K.~N. Solanki, Characterizing the
  local primary dendrite arm spacing in directionally solidified dendritic
  microstructures, Metallurgical and Materials Transactions A 45~(1) (2014)
  426--437.

\bibitem{takaki2016primary}
T.~Takaki, S.~Sakane, M.~Ohno, Y.~Shibuta, T.~Shimokawabe, T.~Aoki, Primary arm
  array during directional solidification of a single-crystal binary alloy:
  large-scale phase-field study, Acta Materialia 118 (2016) 230--243.

\bibitem{liu2020prediction}
S.~Liu, E.~Mart{\'\i}nez, J.~LLorca, Prediction of the {Al-rich part of the
  Al-Cu} phase diagram using cluster expansion and statistical mechanics, Acta
  Materialia 195 (2020) 317--326.

\bibitem{liu2020mgzn}
S.~Liu, G.~Esteban-Manzanares, J.~LLorca, First-principles analysis of
  precipitation in {Mg-Zn} alloys, Physical Review Materials 4 (2020) 093609.

\bibitem{hoyt2003atomistic}
J.~J. Hoyt, M.~Asta, A.~Karma, Atomistic and continuum modeling of dendritic
  solidification, Materials Science and Engineering: R: Reports 41~(6) (2003)
  121--163.

\bibitem{brown2017interfacial}
N.~T. Brown, E.~Martinez, J.~Qu, Interfacial free energy and stiffness of
  aluminum during rapid solidification, Acta Materialia 129 (2017) 83--90.

\bibitem{liu2017multiscale}
H.~Liu, B.~Bell{\'o}n, J.~LLorca, Multiscale modelling of the morphology and
  spatial distribution of $\theta$' precipitates in al-cu alloys, Acta
  Materialia 132 (2017) 611--626.

\bibitem{liu2019precipitation}
H.~Liu, I.~Papadimitriou, F.~Lin, J.~LLorca, Precipitation during high
  temperature aging of al- cu alloys: A multiscale analysis based on first
  principles calculations, Acta Materialia 167 (2019) 121--135.

\end{thebibliography}

\end{document}